\documentclass[pre,amsmath,twocolumn,showpacs]{revtex4}
\usepackage{bm}
 \usepackage{amssymb}
 \usepackage{graphicx}
 \usepackage{subfigure}
\usepackage[dvips]{color}

\begin{document}

 \title{Jamming by Shape in Kinetically-Constrained Models}

 \author{Eial Teomy}
 \email{eialteom@post.tau.ac.il}
 \author{Yair Shokef}
 \email{shokef@tau.ac.il}

 \affiliation{School of Mechanical Engineering, Tel Aviv University, Tel Aviv 69978, Israel}

 \begin{abstract}

We derive expressions for the critical density for jamming in a hyper-rhomboid system of arbitrary shape in any dimension for the Kob-Andersen and Fredrickson-Andersen kinetically-constrained models. We find that changing the system's shape without altering its total volume or particle density may induce jamming. We also find a transition between shapes in which the correlation length between jammed particles is infinite, and shapes that have a finite correlation length.
 
 \end{abstract}


 \maketitle

\section{Introduction}

 Increasing the density of particles in granular materials causes them to undergo a transition from an unjammed state, in which the particles can move relatively freely, to a jammed state, in which almost none of the particles can move~\cite{jamming,vanHecke_Review_2010,Liu_Nagel_Review_2010,Biroli_Garrahan_Perspective_2013}. Systems of interest in nature and in industrial applications typically have complicated geometries which strongly affect jamming in them~\cite{industry,industry2,nature,nature2}, and it is thus important to understand how does confinement influence the jamming of granular matter. Here we investigate the effects of confinement on the jamming transition, and in particular test how does the shape of containers, and not only their volume, determine how they jam. Most theoretical work so far was done on square and cubic systems, or even in the infinite system size limit \cite{durian,haxton,ohern,ningxu,lerner,barrat}.
 
 There are numerous laboratory experiments that deal with non-square two-dimensional systems \cite{exp1,exp2,exp3,exp4}. For example, Daniels and Behringer conducted an experiment on polypropylene spheres in an annulus \cite{behringer1}, which is large enough to be considered a rectangle with infinite length and finite width. A different experiment by Bi et al. \cite{behringer2} consists of shearing a square system such that it becomes a rectangle with the same area and particle density as the original square. Other experiments, such as \cite{qpexp,yunker,antina,cheng}, considered colloids confined between quasiparallel plates.

In this Paper we study the effects of confinement on jamming by studying these phenomena in kinetically-constrained models in d-dimensional hyper-rhomboids. The essence of jamming is captured by the various kinetically-constrained models \cite{review,review2,kronig,fieldings,toninelli,sellitto,knights,DFOT,sellitto2,spiral,spiral3d,jeng,shokef,elmatad,driving2,driving}. Such simple models, and other related models \cite{kipnis,derrida2,derrida,bouchaud,asep,monthus,nemeth,scheidler,varnik,srebro,zrp,lang}, have exact solutions, which help us to obtain deep insights on the physical processes responsible for the corresponding phenomena in real systems.

 \subsection{Models}

We consider the Kob-Andersen (KA) \cite{ka} and Fredrickson-Andersen (FA) \cite{fa,fa2} models on hyper-cubic lattices of arbitrary dimension $d$. The system is confined to a hyper-rhombus of size $L_{1}\times ...\times L_{d}$, such that $L_{i}\leq L_{i+1}$ for all $1\leq i\leq d-1$. In the KA model each site may be occupied by at most one particle, and a particle can move to a neighboring vacant site if it has at least $m$ neighboring vacancies before and after the move. In the FA model a site can change its state from occupied to vacant or vice versa if it has at least $m$ neighboring vacancies. If $m=1$, then even a single vacancy can facilitate the movement of the entire system, and thus all the particles can move eventually regardless of the system's shape or size. If $m$ is less than or equal to the system's dimension, $d$, there are no frozen particles in an infinite system. However, if at least $d-m+1$ of the system's sides are of finite size, there is a finite probability that there will be a frozen structure spanning the system from one edge to the other. If $m$ is larger than the system's dimension, consider a completely occupied $d$-dimensional small hypercube inside the system. Each particle in this hypercube has at most $d$ neighboring vacancies, and so will never move. Hence, there is always a finite fraction of frozen particles even for an infinite system. Moreover, these frozen particles are not necessarily connected to the system's edges and can be ``free-floating" in the middle, and we do not refer to this situation as jammed. In general, there is a finite probability for having frozen particles as long as at least $d-m+1$ of the rhomboid's sides are finite. 

For infinite systems, the critical density is the density at which the fraction of frozen particles is singular \cite{knights,spiral,spiral3d}. Although no real phase transition occurs in finite systems, there is a crossover transition between an almost unfrozen regime, in which almost all of the particles on average are unfrozen, to a frozen regime, in which almost all the particles are frozen. We use the average fraction of frozen particles, $n_{PF}$, as the order parameter, and define the critical density as the density at which on average half of the particles are frozen, $n_{PF}=1/2$. With hard-wall boundary conditions, our order parameter can be thought of as a point-to-set correlation function \cite{pts,pts2}, because it measures the probability that a particle is frozen given that the outside edges of the system are blocked.

Balogh et al showed that for the FA model in a hyper-cube with $L_{i}=L$, the critical vacancy density is \cite{balogh}
\begin{align}
&v_{d,m}=\left(\frac{\lambda_{d,m}}{\ln_{(m-1)} L}\right)^{d-m+1} ,\label{vcgen1}
\end{align}
where $\ln_{(k)}(x)=\ln\left(\ln\left(...\ln(x)...\right)\right)$ is the $\ln$ function iterated $k$ times,
\begin{align}
&\lambda_{d,m}=\int^{\infty}_{0}g_{m-1}(z^{d-m+1})dz ,\label{lamtheo}
\end{align}
and
\begin{align}
&g_{k}(z)=-\ln\left[\frac{1-e^{-kz}+\sqrt{\left(1+e^{-kz}\right)^{2}-4e^{-(k+1)z}}}{2}\right] .
\label{lamtheo2}
\end{align}
In particular, for $d=2$, the only nontrivial case is $m=2$, for which 
\begin{align}
&v_{2,2}=\frac{\lambda_{2,2}}{\ln L},\nonumber\\
&\lambda_{2,2}=\frac{\pi^2}{18}\approx0.5483... ,\label{lam22}
\end{align}
and for $d=3$, both $m=2$ and $m=3$ are nontrivial, with
\begin{align}
\begin{array}{ll}
v_{3,2}=\left(\frac{\lambda_{3,2}}{\ln L}\right)^{2},&v_{3,3}=\frac{\lambda_{3,3}}{\ln\ln L}\\
\lambda_{3,2}=0.9924...,&\lambda_{3,3}=0.4039... .
\end{array}
\end{align}

This is correct only asymptotically for very large systems. For finite systems, there is some effective value of $\lambda_{d,m}$ which converges to these values for extremely large $L$ \cite{slowcon}. In what follows, whenever $\lambda_{d,m}$ appears, it should be understood as this effective $\lambda_{d,m}$.

Although for $m\leq d$ the critical vacancy density in the thermodynamic limit $(L\rightarrow\infty)$ is $v_{c}=0$, it is still important to know the effects that the size and shape of the system have on $v_{c}$, since real and simulated systems are of finite size and can be of different shapes. It is also interesting to investigate systems which are infinite or very large in some directions and finite or small in other directions, such as very long tunnels that have a finite width. In \cite{teomy} we investigated two-dimensional rectangular systems in the $m=2$ FA and KA models, and found that the critical vacancy density in infinitely long tunnels with a finite width $W$ scales as $v_{2,2}\propto\frac{1}{\sqrt{W}}$, which is qualitatively different from $v_{c}\propto1/\ln L$ for square systems in Eq. (\ref{lam22}). Here, we expand our previous work by considering $d$-dimensional hyper-rhomboids with any $1<m\leq d$. 

\subsection{Outline}

In this work we distinguish between weak confinement and strong confinement, depending on the number of effective dimensions of the hyper-rhomboid, $d_{eff}$. A tunnel, for example, has one effective dimension, $d_{eff}=1$, while a 3D system confined between closely separated parallel walls has $d_{eff}=2$. Consider a system in which $d_{eff}$ of its sides are infinite, and the other $s=d-d_{eff}$ sides are finite. If $d_{eff}\geq m$, then none of the particles in the system are frozen, while if $d_{eff}<m$ a finite fraction of the particles is frozen. When the system is of finite size, its critical density is mainly governed by the size of the largest $d_{eff}$ dimensions if $d_{eff}\geq m$, or by the size of the smallest $s$ dimensions if $d_{eff}<m$. Hence, we define a weakly confined system as a system with $d_{eff}\geq m$ (since the confinement in the $s$ small dimensions has only a weak effect), and a strongly confined system as a system with $d_{eff}<m$. 

A weakly confined system is characterized by long range correlations. By making a small, local change in the configuration, the entire system can change from being unjammed to jammed and vice versa. In a strongly confined system, the correlations are short-ranged. There are clusters of frozen and unfrozen particles, such that a small local change in the configuration may change the state of the local cluster, but not of the entire system.

We also find analytical expressions for the critical density for different hyper-rhomboids. By changing the system's shape, but not its total volume or particle density, the critical density also changes, such that it may be above the particle density for some shape but below it for a different shape. Hence, a system may jam and/or unjam by changing only its shape.

The remainder of this article is organized as follows. In Section \ref{SectionCrit} we analyze the critical densities for hyper-rhomboids. We present here a derivation of the critical density for three-dimensional systems, and retain the derivation for general dimensions to Appendix \ref{critd4}. In Section \ref{SectionJamShape} we show how changing the shape of a system induces jamming by studying the $m=2$ model in three dimensional rhomboids. In Section \ref{SectionCorrelation} we study the correlation between frozen sites in strongly confined three-dimensional systems in the $m=3$ model. Section \ref{SectionSummary} summarizes our work.

\section{Critical Density}
\label{SectionCrit}
In this Section we show the derivation of the critical density for $d$-dimensional rhomboids in any $m\leq d$. We first present a sketch of the derivation, and later show the full details. Those not interested in the more technical details can read only the beginning of this section and skip the subsections. A detailed derivation of the critical density for general dimensions appears in Appendix \ref{critd4}.

Consider first a special case of weakly confined systems, namely hyper-cubes. The system is unfrozen if it contains at least one critical droplet, which is a small unblocked region which can cause the entire system to become unblocked \cite{holroyd}. We denote the probability that a site is part of a critical droplet which unblocks a hyper-cube of size $L^d$ by $P\left[L\times...\times L\right]$. The average number of sites which seed a critical droplet is thus $L^{d}P\left[L\times...\times L\right]$. The system is unfrozen when it contains at least one critical droplet. Therefore, the critical density is found by solving the equation
\begin{align}
1=L^{d}P\left[L\times...\times L\right] .\label{eqplr}
\end{align}
Holroyd \cite{holroyd} showed that for two-dimensional systems with $m=2$ the probability to seed a droplet is
\begin{align}
P\left[L\times L\right]=\exp\left[-2\lambda_{2,2}/\left(1-\rho\right)\right] ,
\end{align}
where $\lambda_{2,2}$ is given in Eq. (\ref{lam22}) above. Thus Eq. (\ref{vcgen1}) is retrieved for the two-dimensional square with $m=2$. Balogh et. al. \cite{balogh} used the same idea for $d$-dimensional hyper-cubes in any $m\leq d$ to derive Eq. (\ref{vcgen1}) for hyper-cubes in any dimension.

The main idea is to start from a very small unblocked region and try to expand it isotropically in all directions by checking if the sites adjacent to the unblocked region satisfy some condition (which depends on $d$ and $m$). Therefore, the most probable shape that a droplet can unblock is a hyper-cube. However, other shapes, as long as they are not too deformed, can also be unblocked by a single droplet, because the system's edges do not impede the expansion of the droplet. We show below (for $d=3$) and in Appendix \ref{critd4} (for $d>3$) that for weakly confined systems ($d_{eff}\geq m$), if there is a critical droplet it will unblock the entire system with high probability. The implication of this is that if a particle is frozen, there is no critical droplet, and thus the other particles in the system are highly likely to be frozen as well. Hence, the correlation between frozen particles spans the entire system, no matter its size. By calculating the probability that a particle in a hyper-rhomboid is part of a critical droplet, $P\left[L_{1}\times...\times L_{d}\right]$, and solving the analog to Eq. (\ref{eqplr})
\begin{align}
1=VP\left[L_{1}\times...\times L_{d}\right] ,
\end{align}
where $V$ is the volume of the system, we find the critical vacancy density
\begin{align}
v^{weak}_{c}=\frac{1}{V_{s}}\left(\frac{\lambda_{d_{eff},m}}{\ln_{(m-1)}V^{1/d_{eff}}}\right)^{d_{eff}-m+1} ,
\end{align}
where $s=d-d_{eff}$ and
\begin{align}
V_{s}=\prod^{s}_{i=1}L_{i} ,
\end{align}
is the volume associated with the small dimensions. Except for the prefactor of $1/V_{s}$, the critical density is the same as for a hyper-cube with $d_{eff}$ dimensions. Note that for weakly confined systems the critical density depends on the ratios between the lengths $L_{i}$ only via the number of effective dimension. This means that as long as the hyper-rhomboid is not too deformed, its critical density is equal to the critical density of a hyper-cube of equal volume.

As will be explained in subsections \ref{critm2} and \ref{critm3}, due to the way the droplet is constructed and expanded, it needs at least $m$ dimensions in order to expand indefinitely. Therefore, in a strongly confined system $(d_{eff}<m)$, the system's edges block the expansion of the critical droplet, and thus each droplet unblocks only part of the system. These unfrozen clusters have a characteristic size $\xi$, and we assume that at the critical density the frozen and unfrozen clusters have the same characteristic size. The critical density is found by evaluating the density at which half of the clusters contain a droplet, i.e. by solving the equation
\begin{align}
\frac{1}{2}=V_{cluster}P\left[cluster\right] ,
\end{align}
where $V_{cluster}$ is the volume of a typical cluster and $P\left[cluster\right]$ is the probability that a site will seed a critical droplet that unblocks the cluster.
The solution to this equation is
\begin{align}
v^{strong}_{c}\sim\left[\ln_{(m-2)}V^{1/(d-t)}_{s}\right]^{-d+t+m-1} ,
\end{align}
where $t$ is the number of very small dimensions (for example, a 3D system with $L_{3}\gg L_{2}\gg L_{1}$ has $s=2$ and $t=1$), and $\ln_{(0)}x\equiv x$. This expression is similar to the critical density of a weakly confined system with $d-t-1$ dimensions in a $m-1$ model. In effect, the strong confinement removes one degree of freedom.

It is important to note that the value of the parameter $\lambda$ given in Eq. (\ref{lamtheo}) is valid only for very large systems. For $m=2$ in $d=2$ it is well known that even for the largest systems that have been simulated ($L\approx10^5$) $\lambda_{2,2}$ is equal to roughly half of its asymptotic value \cite{slowcon,lambda22}. By considering other ways to expand the droplet, Holroyd \cite{slowcon} evaluated the asymptotic corrections to this value, and estimated that systems as large as $L=10^{20}$ would be needed to observe convergence in the numerical value of $\lambda_{2,2}$. Here we see a similar behavior for $d=3$.

\begin{figure}
\includegraphics[width=\columnwidth]{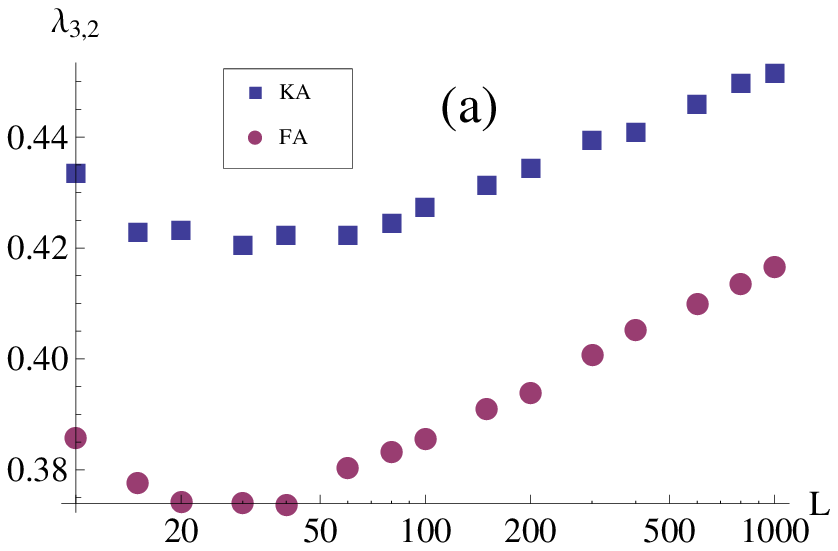}\\
\includegraphics[width=\columnwidth]{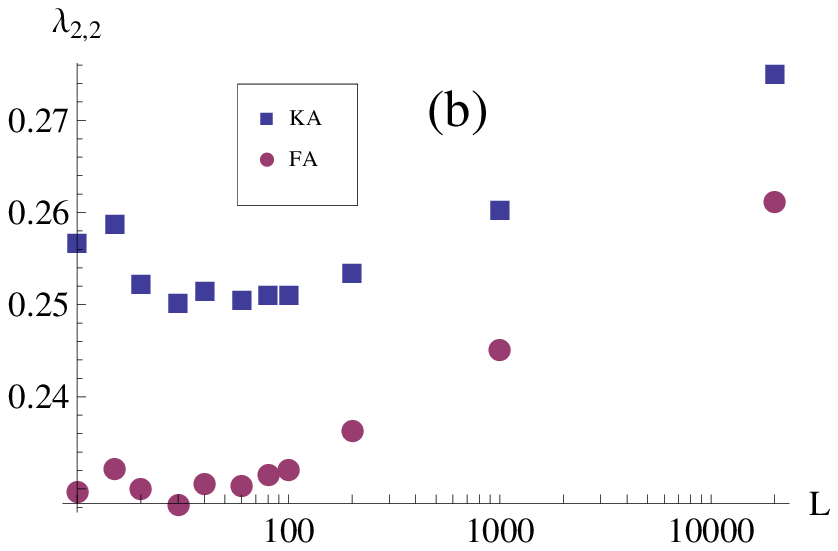}
\caption{The effective values of $\lambda_{3,2}=\sqrt{v_{c}}\ln L$ for cubes of size $L^{3}$ (a) and of $\lambda_{2,2}=v_{c}\ln L$ for squares of size $L^2$ (b) in the KA (squares) and the FA (circles) models. The values of $\lambda$ in the sizes sampled are far from their asymptotic values of $\lambda_{3,2}\approx0.99$ and $\lambda_{2,2}\approx0.54$, and they appear to behave as $\sim\ln L$.}
\label{l32eff}
\end{figure}

\begin{figure}
\includegraphics[width=\columnwidth]{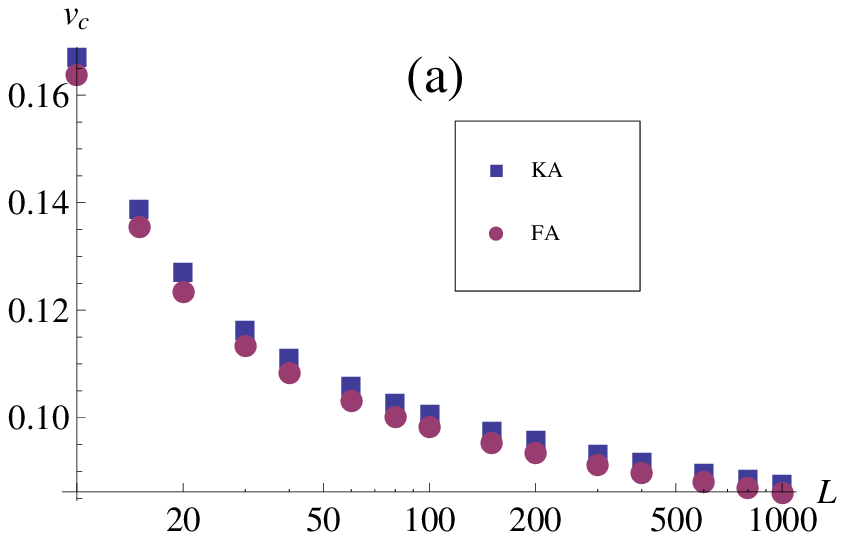}\\
\includegraphics[width=\columnwidth]{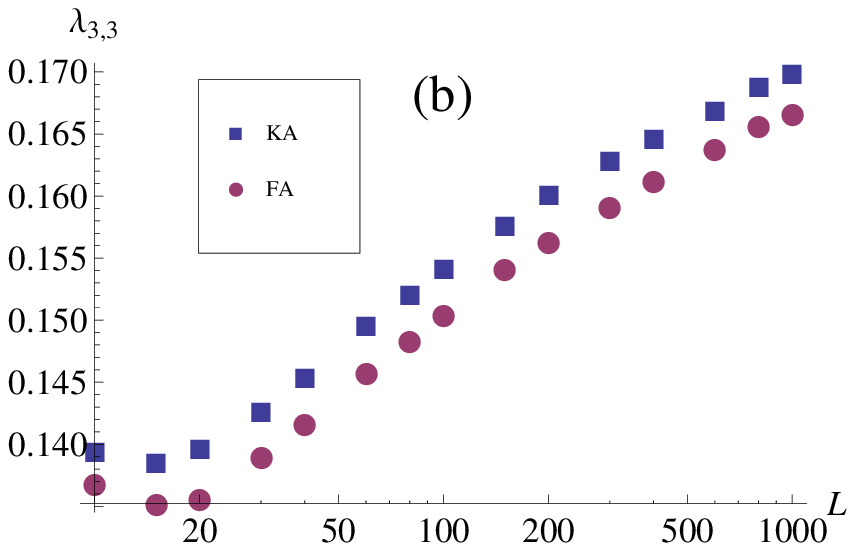}
\caption{The critical vacancy density $v_{c}$ (a) and the effective $\lambda_{3,3}=v_{c}\ln\ln L$ (b) for three-dimensional cubes as a function of the cube size $L$ in the $m=3$ model. The effective $\lambda_{3,3}$ is far from its asymptotic value of $0.4$}
\label{vcm3}
\end{figure}

Figure \ref{l32eff} shows the effective value of $\lambda_{3,2}=\sqrt{v_{c}}\ln L$ in cubes of size $L^{3}$ and of $\lambda_{2,2}=v_{c}\ln L$ in squares of size $L^{2}$. Under our current computational limits ($L^d\leq10^{9}$, which requires $30GB$ memory), the effective $\lambda_{3,2}$ has a value of approximately $0.45$ in the KA model and $0.42$ in the FA model, far from its asymptotic value of $\approx0.99$. However, it is clear that it grows with $L$. Theoretically, the value of $\lambda$ for both models should be the same. The added constraint in the KA model reduces the probability to unblock the entire system, and thus increases $\lambda$, and this effect is non-negligible in small systems. Figure \ref{vcm3} shows the critical vacancy density and the effective $\lambda_{3,3}$ for cubes of different sizes in the $m=3$ model. Similarly to the $m=2$ model, the effective $\lambda$ is far from its asymptotic value of $0.4$.

\subsection{Derivation of the critical density for the $m=2$ model}
\label{critm2}

In this subsection we first sketch the derivation of Eq. (\ref{vcgen1}) given in \cite{balogh} for hypercubes in arbitrary dimensions for the $m=2$ model. For simplicity we give it here for three-dimensions. The generalization to higher dimensions is straightforward. We will then consider three-dimensional rhomboids.

\subsubsection{Cubes}

When $m=2$, a cube $(L_{1}=L_{2}=L_{3}\equiv L)$ is either completely unfrozen or almost completely frozen. The system can become unfrozen if it contains at least one critical droplet. We denote the probability that a site is part of a critical droplet which unblocks a system of size $L\times L\times L$ by $P\left[L\times L\times L\right]$. The average number of sites which seed a critical droplet is thus $L^{3}P\left[L\times L\times L\right]$. The system is unfrozen when it contains at least one critical droplet. Therefore, the critical density is found by solving the equation
\begin{align}
1=L^{3}P\left[L\times L\times L\right] .\label{l3peq1}
\end{align}

\begin{figure}
\subfigure[]{\includegraphics[width=30pt]{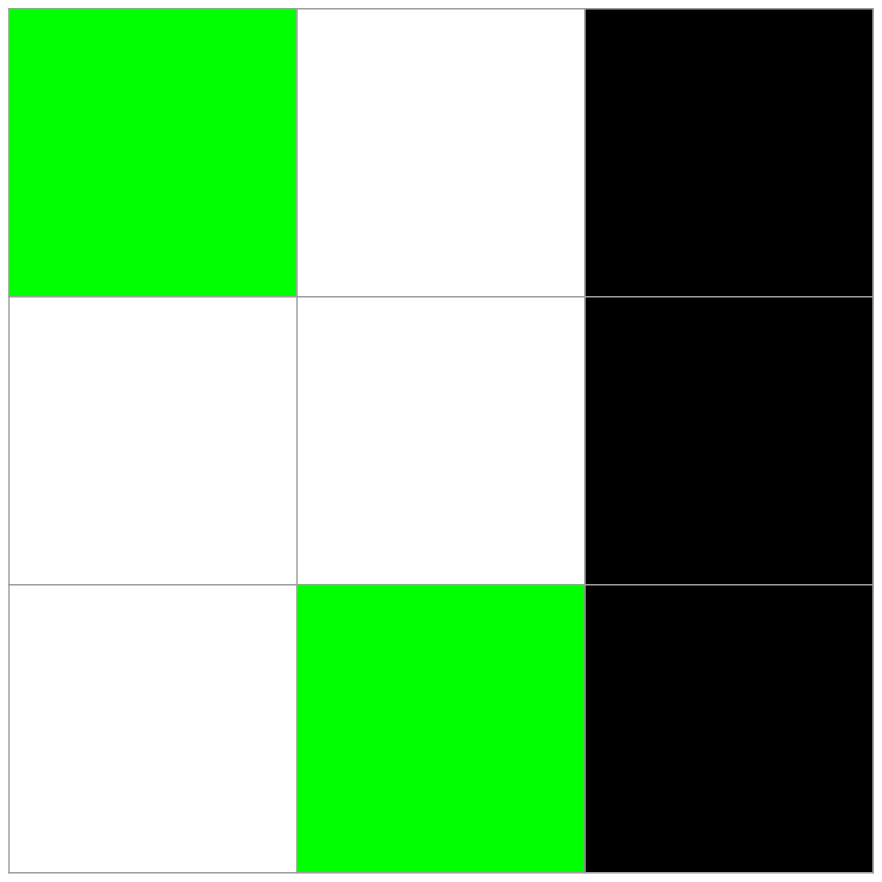}}
\subfigure[]{\includegraphics[width=40pt]{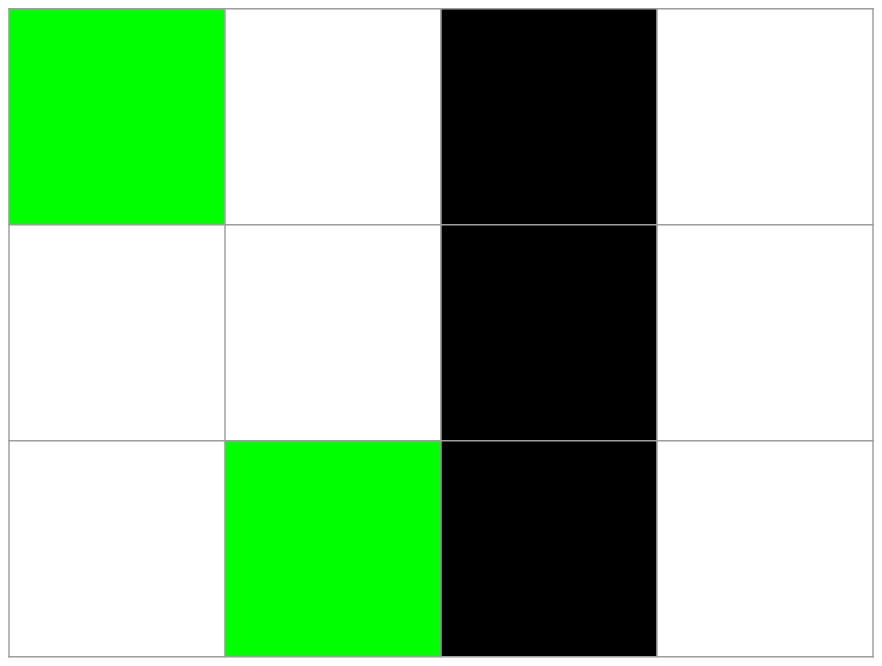}}
\subfigure[]{\includegraphics[width=40pt]{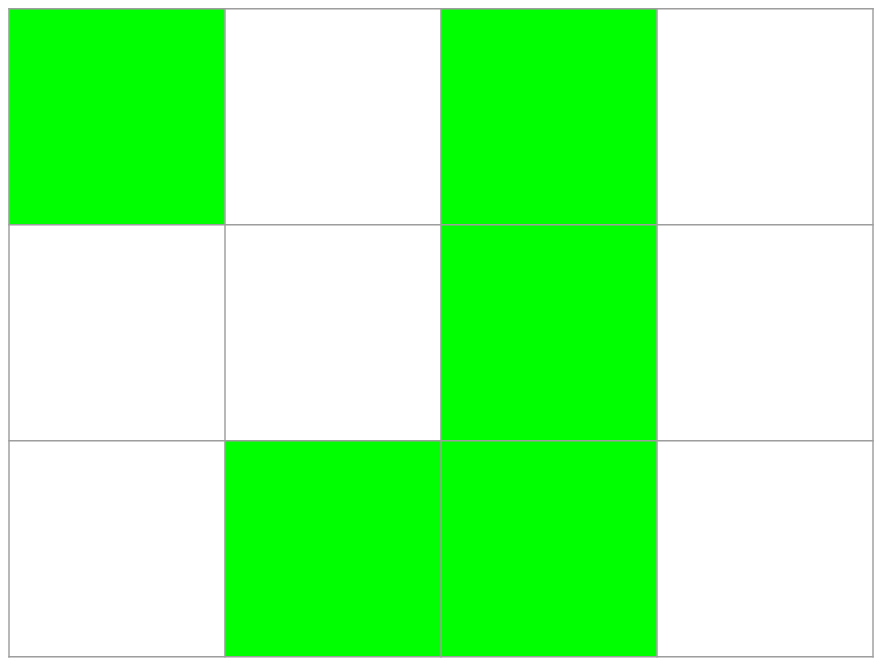}}
\subfigure[]{\includegraphics[width=40pt]{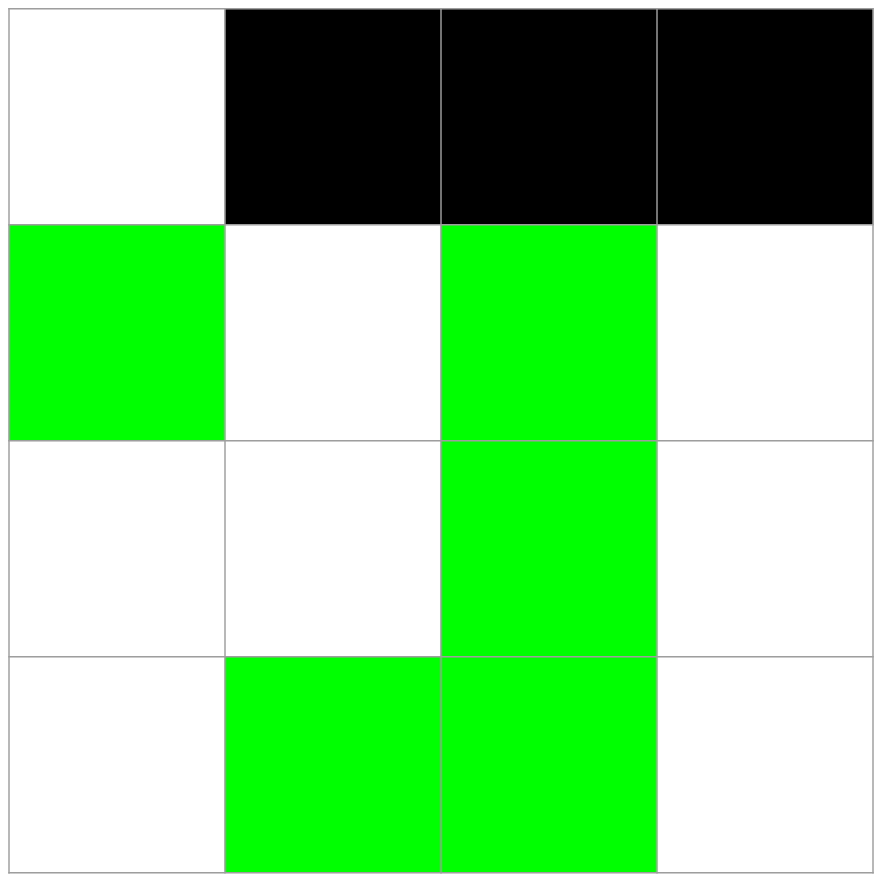}}
\subfigure[]{\includegraphics[width=40pt]{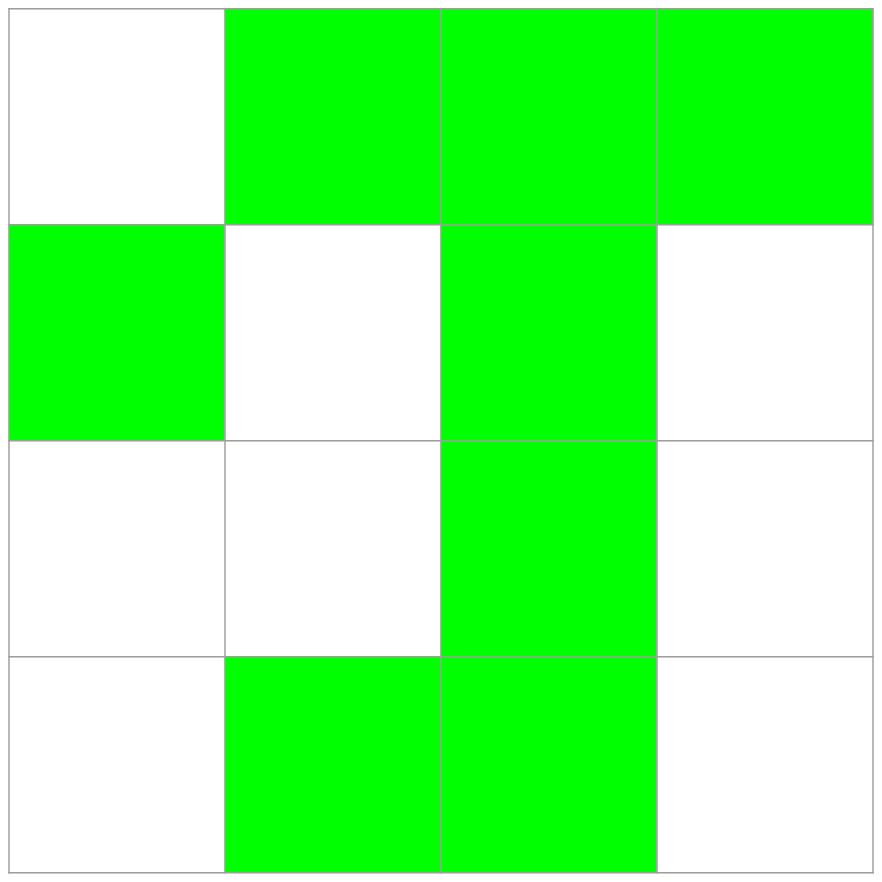}}
\caption{An example of the expansion of a critical droplet in $d=2$. We start from panel (a) with a $3\times 3$ system. The particles shown in green are unfrozen, and the black particles are those that are either blocked at that stage of the expansion or that are newly considered.
 Now, expand it to the right (b). The rightmost column is not full, and the $3 \times 2$ rectangle is unfrozen. This means that the $3 \times 3$ square satisfies the second condition, and the $3 \times 4$ rectangle is unfrozen (c). We now expand it one row upwards (d). The top row is not full, which means that the first condition is satisfied, and thus the entire $4 \times 4$ square is unfrozen (e).}
\label{cubeex}
\end{figure}

In order for a site to be in a critical droplet, we first consider a small $2\times2\times2$ cube and randomly choose one of the eight $3\times3\times3$ cubes that contain it. We then check whether this $3\times3\times3$ cube is frozen or not. We want to continue this process to a $4\times4\times4$ cube and so on until the entire system is unfrozen. The original $2\times2\times2$ cube is a critical droplet only if it is possible to expand it to the system's size. Assume that we now want to check whether a box of size $(l+1)\times(l+1)\times(l+1)$ is frozen or not. The probability that it is not frozen is $P\left[(l+1)\times(l+1)\times(l+1)\right]$. We start from a cube of size $l\times l\times l$ and expand it in three steps: first to a rhomboid of size $l\times l\times (l+1)$ then to $l\times(l+1)\times(l+1)$ and lastly to $(l+1)\times(l+1)\times(l+1)$. In the first step, the rhomboid is unblocked if one of two disjoint conditions occur: either the $l\times l\times l$ box is unblocked and the added side is not full, or the $l\times l\times (l-1)$ rhomboid is unblocked, the sites in the side of the $l\times l\times l$ box but not in the $l\times l\times (l-1)$ rhomboid are all occupied and the added side of the $l\times l\times (l+1)$ rhomboid is not full. See Fig. \ref{cubeex} for an illustration of the expansion process in two-dimensions. Hence $P\left[l\times l\times(l+1)\right]$ satisfies the equation
\begin{align}
&P\left[l\times l\times(l+1)\right]=P\left[l\times l\times l\right]\left(1-\rho^{l^{2}}\right)+\nonumber\\
&+P\left[l\times l\times(l-1)\right]\rho^{l^{2}}\left(1-\rho^{l^{2}}\right) ,\label{prec}
\end{align}
where $\rho$ is the particle density.

Assuming that these probabilities have the form
\begin{align}
&P\left[l_{1}\times l_{2}\times l_{3}\right]=\nonumber\\
&=\prod^{l_{1}}_{l=1}\beta(l^{2})\prod^{l_{2}}_{l=1}\beta(l^{2})\prod^{l_{3}}_{l=1}\beta(l^{2}) ,\label{pform}
\end{align}
yields 
\begin{align}
&P\left[l\times l\times(l+1)\right]=\nonumber\\
&=P\left[l\times l\times(l-1)\right]\beta\left[(l+1)^2\right]\beta(l^2) ,
\end{align}
and
\begin{align}
&P\left[l\times l\times l\right]=P\left[l\times l\times(l-1)\right]\beta(l^2) .
\end{align}
Using this assumption in Eq. (\ref{prec}) yields
\begin{align}
&\beta[(l+1)^{2}]\beta(l^{2})=\left(\beta(l^{2})+\rho^{l^{2}}\right)\left(1-\rho^{l^{2}}\right) .\label{beq}
\end{align}
Further assuming that for large $l$, $\beta$ depends weakly on it, namely $\beta\left[(l+1)^{2}\right]\approx\beta\left(l^{2}\right)$ in Eq. (\ref{beq}), we have a quadratic equation for $\beta$ with the solution
\begin{align}
\beta(l^{2})=\frac{1-\rho^{l^{2}}+\sqrt{1+2\rho^{l^{2}}-3\rho^{2l^{2}}}}{2} .\label{betaf}
\end{align}

The other two steps of the rhomboid expansion yield similar equations for $\beta$. By assuming that $\beta$ depends weakly on $l$ for large $l$, the solution of these two equations is the same as in Eq. (\ref{betaf}).

\begin{figure}
\subfigure{\includegraphics[width=\columnwidth]{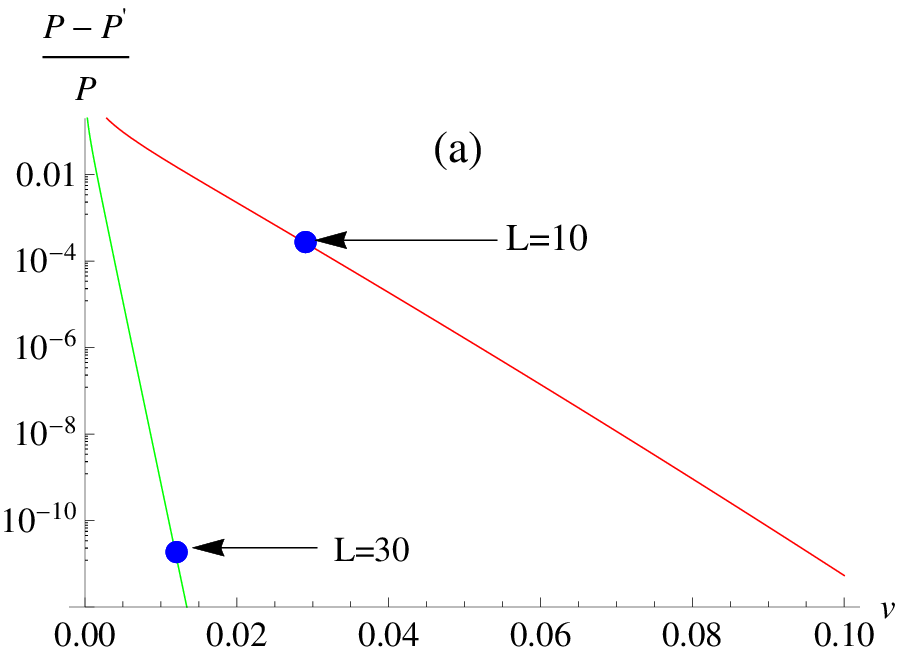}}\\
\subfigure{\includegraphics[width=\columnwidth]{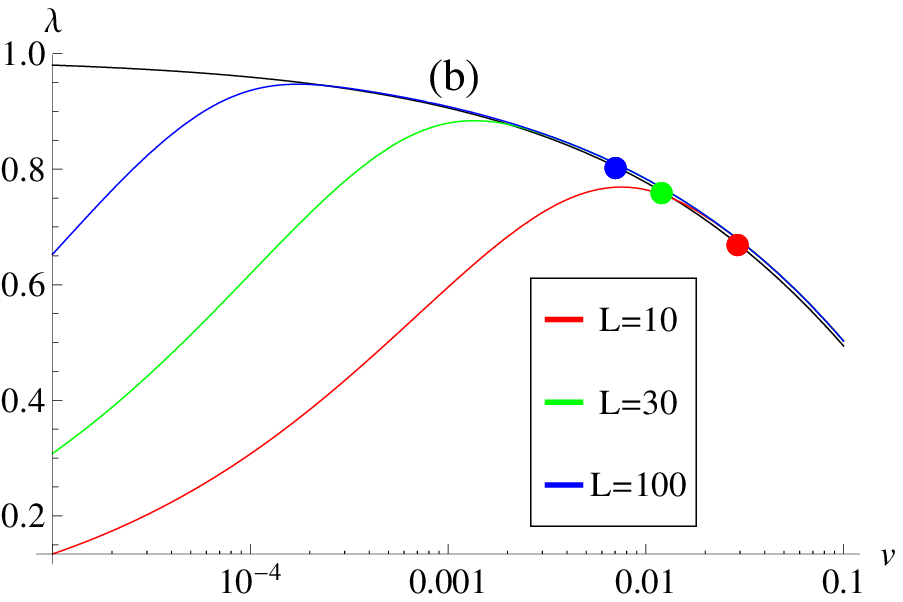}}
\caption{(a) The relative difference between $P'=P\left[1000\times1000\times1000\right]$ and $P=P\left[L\times L\times L\right]$ for $L=10$ (red) and $L=30$ (green) as a function of the vacancy density $v$, and (b) A comparison between the sum $-\sqrt{v}\sum^{L}_{l=1}\ln\beta(l^{2})=-\frac{\sqrt{v}}{3}\ln P\left[L\times L\times L\right]$ and the integral in Eq. (\ref{lamtheo32}), $\lambda_{3,2}+\sqrt{v}\left(\frac{\ln v}{4}-1\right)$, as a function of the vacancy density $v$ for different sizes: $L=10$ (red), $L=30$ (green), and $L=100$ (blue). The black line is $\lambda_{3,2}+\sqrt{v}\left(\frac{\ln v}{4}-1\right)$. The dots mark the critical density for $L=10$ $(v_{c}\approx0.029)$, $L=30$ $(v_{c}\approx0.012)$ and $L=100$ $(v_{c}\approx0.007)$. The relative difference is very small at the critical vacancy density ($v_{c}\approx0.029$ for $L=10$ at which the relative difference is $0.0003$ and $v_{c}\approx0.012$ for $L=30$ at which the relative difference is $3\times10^{-10}$). For a given $L$, $\lambda_{eff}$ is very close to $\lambda_{3,2}+\sqrt{v}\left(\frac{\ln v}{4}-1\right)$ until a certain value of the density where it drops rapidly. It is clear that for each $L$, around $v_{c}$ the approximation is valid.}
\label{prel1}
\end{figure}

Therefore, the probability that a certain site is part of a critical droplet, i.e. that it can unblock the entire system is
\begin{align}
P\left[L\times L\times L\right]=\left[\prod^{L}_{l=1}\beta(l^2)\right]^{3}=\exp\left[3\sum^{L}_{l=1}\ln\beta(l^{2})\right] .\label{piii}
\end{align}
Since the critical density in an infinite system is $1$ and the critical density in finite systems is close to $1$ we can approximate $\rho^{l^{2}}$ by
\begin{align}
\rho^{l^{2}}\approx\exp\left[-vl^{2}\right] ,
\end{align}
where $v=1-\rho$ is the vacancy density. For large enough $L$ ($L\geq10$), the value of the probability $P\left[L\times L\times L\right]$, calculated via Eq. (\ref{piii}), is very close to $P\left[\infty\times\infty\times\infty\right]$ as shown in Fig. \ref{prel1}, so we can use the latter even for finite systems.

Changing the summation in Eq. (\ref{piii}) to an integration over $z=\sqrt{v}l$, yields according to the Euler-Maclaurin formula
\begin{align}
&P\left[\infty\times\infty\times\infty\right]\approx\exp\left\{3\int^{\infty}_{\sqrt{v}}\ln\left[\beta\left(\frac{z^{2}}{v}\right)\right]\frac{dz}{\sqrt{v}}+\right.\nonumber\\
&\left.+\frac{3}{2}\left[\ln\beta(\infty)+\ln\beta(1)\right]\right\}=\nonumber\\
&=\exp\left\{3\int^{\infty}_{\sqrt{v}}\ln\left[\beta\left(\frac{z^{2}}{v}\right)\right]\frac{dz}{\sqrt{v}}+\frac{3}{2}\ln\beta(1)\right\} .
\end{align}
Taking the limit $v\ll1$ yields
\begin{align}
P\left[\infty\times\infty\times\infty\right]=\exp\left\{-\frac{3\left[\lambda_{3,2}+\sqrt{v}\left(\frac{\ln v}{4}-1\right)\right]}{\sqrt{v}}\right\} ,
\end{align}
where
\begin{align}
&\lambda_{3,2}=-\int^{\infty}_{0}\ln\left(\frac{1-e^{-z^{2}}+\sqrt{1+2e^{-z^{2}}-3e^{-2z^{2}}}}{2}\right)\approx\nonumber\\
&\approx0.9924 .\label{lamtheo32}
\end{align}
Using this result in Eq. (\ref{l3peq1}) yields Eq. (\ref{lamtheo}). The transition from the sum to the integral is valid for very small $v$ and large $L$ as shown in Fig. \ref{prel1}.

\subsubsection{Three-dimensional rhomboids - weak confinement}
We now assume that the three length scales $L_{1}\leq L_{2}\leq L_{3}$ are not necessarily equal. 

We repeat the derivation of the critical droplets, such that the critical density is found from the relation
\begin{align}
1=L_{1}L_{2}L_{3}P\left[L_{1}\times L_{2}\times L_{3}\right] .\label{vceq}
\end{align}
We assume that the form of $P$ is
\begin{align}
&P\left[L_{1}\times L_{2}\times L_{3}\right]=\nonumber\\
&=\prod^{L_{1}}_{l=1}\beta^{3}(l^{2})\prod^{L_{2}}_{L_{1}+1}\beta^{2}(L_{1}l)\prod^{L_{3}}_{L_{2}+1}\beta(L_{1}L_{2}) .\label{pform2}
\end{align}
The derivation of the expansion of the critical droplets is the same as in cubes until $l=L_{1}$. At that point, we can only expand the rhomboid in two directions and the size of the third side is fixed to be $L_{1}$. Then, when $l=L_{2}$ we can only expand in one direction and the size of the other two sides are $L_{1}$ and $L_{2}$. Therefore, when $l\leq L_{1}$, $P$ satisfies the same equation as in a cube, Eq. (\ref{prec}), and the function $\beta$ is given by Eq. (\ref{betaf}). When $L_{1}<l\leq L_{2}$ and $L_{2}\leq l\leq L_{3}$, the probability $P$ satisfies similar recursion relations, such that in all three cases 
\begin{align}
\beta(x)=\frac{1-\rho^{x}+\sqrt{1+2\rho^{x}-3\rho^{2x}}}{2} .\label{betadef2}
\end{align}
In order to find the critical density we use Eqs. (\ref{pform2}) and (\ref{betadef2}) in Eq. (\ref{vceq})
\begin{align}
&0=\ln V+3\sum^{L_{1}}_{l=1}\ln\beta(l^2)+2\sum^{L_{2}}_{l=L_{1}+1}\ln\beta(L_{1}l)+\nonumber\\
&+\sum^{L_{3}}_{l=L_{2}+1}\ln\beta(L_{1}L_{2}) ,
\end{align}
where $V=L_{1}L_{2}L_{3}$ is the volume of the system. We now change the first two sums to integrals over $z=\sqrt{v}l$ and $z=vL_{1}l$ respectively and calculate the third sum explicitly
\begin{align}
&0=\ln V-\frac{3}{\sqrt{v}}\int^{\sqrt{v}L_{1}}_{\sqrt{v}}g_{1}(z^{2})dz-\frac{2}{vL_{1}}\int^{vL_{1}L_{2}}_{vL_{1}\left(L_{1}+1\right)}g_{1}(z)dz\nonumber\\
&+\left(L_{3}-L_{2}\right)\ln\beta(L_{1}L_{2}) .
\end{align}
We assume that $v$ is sufficiently small such that the lower limit of the first integral can be taken to zero. If $L_{3}$ is not too large compared to $L_{1}$ and $L_{2}$, we can neglect the third term. Assuming again that $L_{1}$ and $L_{2}$ are large enough this last condition can be written as
\begin{align}
L_{3}\ll\ln^{-1}\beta(L_{1}L_{2})\approx\exp\left[2vL_{1}L_{2}\right] .
\end{align}
For now, we assume that it is satisfied, and thus implicitly assume that $L_{2}\approx L_{3}$, i.e. that the system has at least $d_{eff}\geq2=m$ and so is weakly confined. The case when this condition is not satisfied, and thus the system is strongly confined, is dealt with in the next subsection.

We now look at $vL^{2}_{1}$ and $vL_{1}L_{2}$. If $L_{1}$ is large enough such that $vL^{2}_{1}\gg1$, the system behaves as a cubic system, since $L_{1}$ is large enough to be considered similar to $L_{2}$. If $vL^{2}_{1}$ is small but $vL_{1}L_{2}$ is large, then the system behaves as a quasi-planar system. The third option that $vL_{1}L_{2}\ll1$, cannot exist near the critical density under the assumption that $L_{2}\approx L_{3}$. 

If $\sqrt{v}L_{1}\gg1$, the upper limit of the first integral can be taken to be $\infty$ and the second integral can be neglected. Thus, the equation for the critical density is
\begin{align}
0=\ln V-\frac{3\lambda_{3,2}}{\sqrt{v}} ,
\end{align}
and the solution is
\begin{align}
v^{bulk}_{c}=\left(\frac{3\lambda_{3,2}}{\ln V}\right)^{2} .\label{vcbulk}
\end{align}
Note that in a cube $V=L^{3}$, and so we retrieve Eq. (\ref{vcgen1}) for $m=2$ in three-dimensional systems. This result means that in the bulk regime, the ratio between the different length scales is unimportant, and the only relevant quantity is the system's volume. 

If $\sqrt{v}L_{1}\ll1$, the first integral can be neglected and the limits of the second integral can be taken to be $0$ and $\infty$
\begin{align}
0=\ln V-\frac{2\lambda_{2,2}}{vL_{1}} ,
\end{align}
such that the critical density is
\begin{align}
v^{QP}_{c}=\frac{2\lambda_{2,2}}{L_{1}\ln V} ,\label{vcqp}
\end{align}
which is similar to the behavior of a bulk two-dimensional system
\begin{align}
v^{2D}_{c}=\frac{2\lambda_{2,2}}{\ln V} .
\end{align}

We will now consider what happens when $L_{3}$ is very large, such that $L_{3}\gg\exp\left[2vL_{1}L_{2}\right]$.

\subsubsection{Strong confinement}
When $L_{3}\gg\exp\left[2vL_{1}L_{2}\right]$, we cannot use the notion of critical droplets as is, but must first note that the tunnel is divided into independent sections, similarly to two-dimensional systems \cite{teomy}. We note that if two or more adjacent planes lying in the short directions are completely occupied, then they are permanently frozen. We call $w$ ($w\geq2$) adjacent full planes a \textit{wall} of width $w$. Between each two walls there are a number of planes, which do not contain two or more adjacent occupied planes, and we call these planes a \textit{section} of length $l$. Each section is independent of the others, i.e. its state (whether it is frozen or not) is independent of the configuration of the other sections. We assume that each section behaves as a bulk system of size $L_{1}\times L_{2}\times l$, and take the average value of $l$ as a representative for the whole system. Using combinatorial calculations very similar to those detailed in \cite{teomy}, we find that the relative probability to find a section of length $l$ is
\begin{align}
&Q(l)=\frac{1-\rho^{L_{1}L_{2}}}{\sqrt{1+2\rho^{L_{1}L_{2}}-3\rho^{2L_{1}L_{2}}}}\left[\beta^{l}_{+}-\beta^{l}_{-}\right] ,
\end{align}
with
\begin{align}
&\beta_{\pm}=\frac{1-\rho^{L_{1}L_{2}}\pm\sqrt{1+2\rho^{L_{1}L_{2}}-3\rho^{2L_{1}L_{2}}}}{2} .
\end{align}
The average section length in the $L_{3}$ direction is 
\begin{align}
&\left\langle l_{3}\right\rangle=\frac{\sum^{\infty}_{l=1}Q(l)l}{\sum^{\infty}_{l=1}Q(l)}=\rho^{-2L_{1}L_{2}}+\rho^{-L_{1}L_{2}}-1\approx\nonumber\\
&\approx\rho^{-2L_{1}L_{2}}\approx\exp\left[2vL_{1}L_{2}\right] .
\end{align}
Hence, the division into sections is valid if $L_{3}>\left\langle l_{3}\right\rangle$.

We can now consider a system of size $L_{1}\times L_{2}\times\left\langle l_{3}\right\rangle$, which behaves as either a bulk system or a quasi-planar system. If it behaves as a bulk system, i.e. $\sqrt{v}L_{1}\gg1$, the critical density satisfies the equation
\begin{align}
v^{tunnel}_{c}=\left(\frac{3\lambda_{3,2}}{\ln\left(L_{1}L_{2}\left\langle l_{3}\right\rangle\right)}\right)^{2} ,
\end{align}
which is equivalent to the following cubic equation on $\sqrt{v_{c}}$
\begin{align}
\ln\left(L_{1}L_{2}\right)\sqrt{v_{c}}+2L_{1}L_{2}v^{3/2}_{c}-3\lambda_{3,2}=0 .
\end{align}
In the limit $L_{1}L_{2}\gg\left[\ln\left(L_{1}L_{2}\right)\right]^{3}$, the solution to this equation is
\begin{align}
v^{tunnel}_{c}=\left(\frac{3\lambda_{3,2}}{2L_{1}L_{2}}\right)^{2/3} .\label{vctunnel}
\end{align}
If the section behaves as a quasi-plane, i.e. $\sqrt{v}L_{1}\ll1$, the system is a quasi-tunnel (QT), and the critical density satisfies the equation
\begin{align}
v^{QT}_{c}=\frac{2\lambda_{2,2}}{L_{1}\ln \left(L_{1}L_{2}\left\langle l_{3}\right\rangle\right)} ,
\end{align}
and the solution is
\begin{align}
&v^{QT}_{c}=\frac{\sqrt{16L^{2}_{1}L_{2}\lambda_{2,2}+L^{2}_{1}\ln^{2}\left(L_{1}L_{2}\right)}-L_{1}\ln\left(L_{1}L_{2}\right)}{4L^{2}_{1}L_{2}}\approx\nonumber\\
&\approx\sqrt{\frac{\lambda_{2,2}}{L^{2}_{1}L_{2}}} .\label{vcqt}
\end{align}

If $\left\langle l_{3}\right\rangle>L_{2}$ then there are no more divisions, since the section is built such that it is not divided further. However, if $\left\langle l_{3}\right\rangle<L_{2}$, then a priori it may be that $L_{2}$ is so large that the section is again divided into subsections in the $L_{2}$ direction similarly to what was done before. In Appendix \ref{nodivisions} we show that in fact the section is not divided into smaller subsections.

\subsection{Critical density in the $m=3$ model}
\label{critm3}
We now turn to the $m=3$ model, which is always jammed for $d=2$ but not for $d\geq3$.
We first re-derive the critical density for 3D cubes of size $L\times L\times L$ \cite{balogh}, again using the notion of critical droplets.

\begin{figure}
\subfigure[Top layer]{\includegraphics[width=80pt]{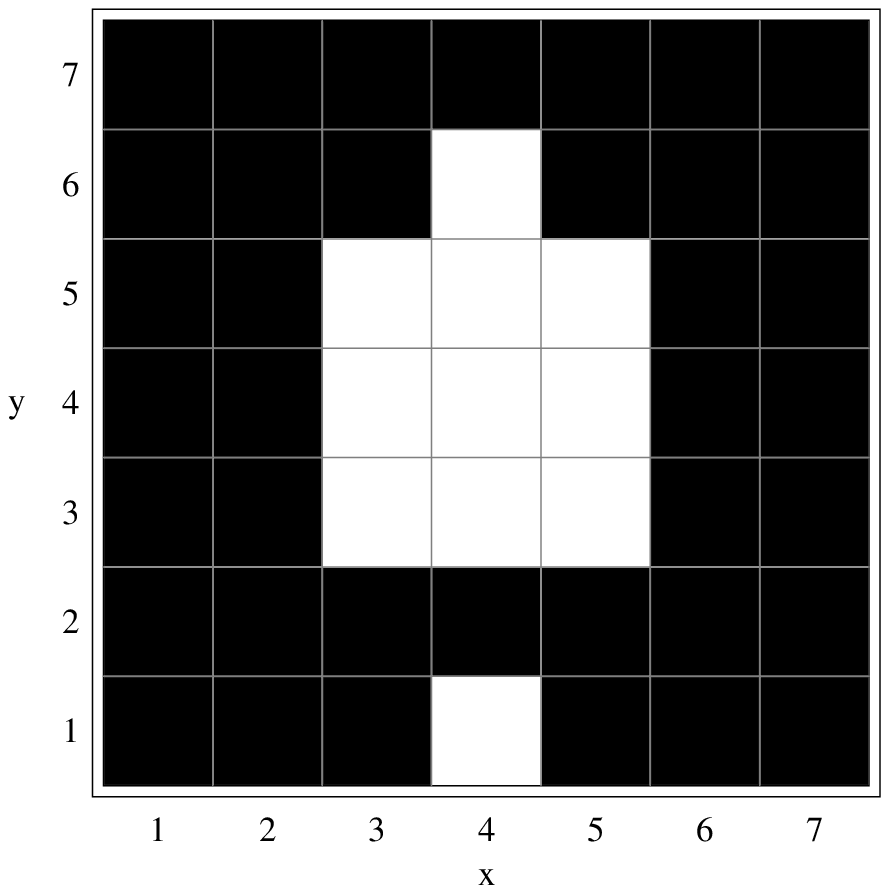}}
\subfigure[Bottom layer]{\includegraphics[width=80pt]{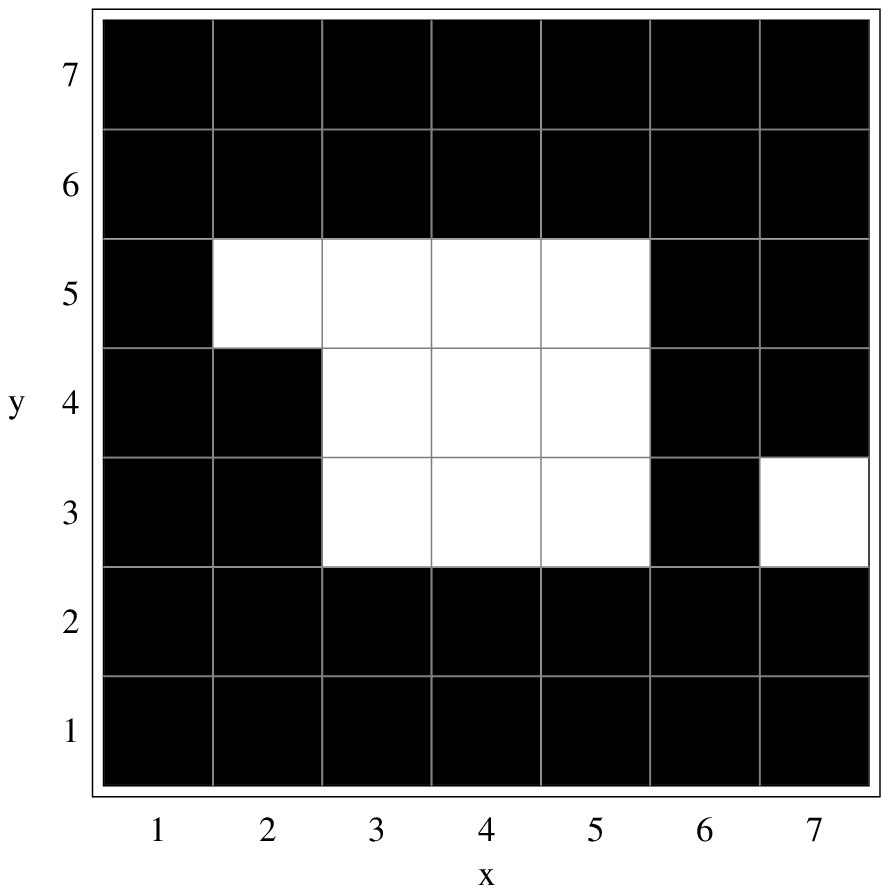}}
\caption{An example for the two layers above the empty cube. Black squares are occupied and white squares are empty. The $3\times3\times2$ rhombus depicted here can be expanded in three of the four possible directions. The rhombus can be expanded in the positive $y$ direction, because its top layer has an adjacent vacancy in that direction in site $(4,6)$. It can be expanded in the positive $x$ direction, because its bottom layer has a vacancy in the next to nearest column in site $(7,3)$. It can be expanded in the negative $x$ direction, because its bottom layer has a vacancy adjacent to it in site $(2,5)$. It cannot be expanded in the negative $y$ direction, because its top layer does not have an adjacent vacancy, it only has a vacancy in the next nearest row at site $(4,1)$, which is not enough.}
\label{3dlayers}
\end{figure}

Consider an unblocked cube of size $L\times L\times L$. The cube can be expanded if the layers adjacent to its sides satisfy some condition which allows them to be unblocked. To construct these conditions, consider a system composed of two 2D layers, each layer of size $L\times L$. These layers correspond to the layers adjacent to one side of the cube. A particle on the bottom layer (the one closer to the original cube) can become unblocked if it has at least two neighboring vacancies in the two layers, and a particle on the top layer can be become unblocked if it has at least three neighboring vacancies. This is similar to the 2D case with $m=2$, so we again use the notion of a critical droplet inside the two layers. An empty rhombus of size $l\times l\times 2$ can be expanded in one direction if one of two conditions is satisfied, similarly to the $m=2$ case: 1) At least one site in the two rows (the one on the top and the one on the bottom) adjacent to the square is empty, or 2) At least one site in the next to nearest bottom row is empty. Figure \ref{3dlayers} illustrates these conditions. 

Therefore, assuming that the probability of a rhombus of size $L\times L\times 2$ to be emptyable is
\begin{align}
P_{2}(L\times L)=\prod^{L}_{l=1}\beta^{2}_{3}(l) ,
\end{align}
we find that the function $\beta_{3}(l)$ satisfies the recursion relation
\begin{align}
\beta^{2}_{3}(l)=\left(1-\rho^{2l}\right)\beta_{3}(l)+\rho^{2l}\left(1-\rho^{l}\right) ,
\end{align}
and is thus equal to
\begin{align}
\beta_{3}(l)=\frac{1-\rho^{2l}+\sqrt{\left(1+\rho^{2l}\right)^{2}-4\rho^{3l}}}{2} .\label{b3}
\end{align}

The face of the cube can be expanded if at least one of the sites on the two layers adjacent to it is a critical droplet. The entire cube can be expanded if it can be expanded in all three directions. Hence, the probability that a cube can be expanded to size $L\times L\times L$ is
\begin{align}
&P_{3}(L\times L\times L)=\prod^{L}_{l=1}\left[1-\left(1-\prod^{l}_{k=1}\beta^{2}_{3}(k)\right)^{2l^{2}}\right]^{3}=\nonumber\\
&=\exp\left\{3\sum^{L}_{l=1}\ln\left[1-\left(1-\exp\left[2\sum^{l}_{k=1}\ln\beta_{3}(k)\right]\right)^{2l^{2}}\right]\right\} .\label{3dexp}
\end{align}
Changing the sums to integrals and taking the upper limit to $\infty$ yields
\begin{align}
P_{3}(L\times L\times L)\approx\exp\left[-3\mu\exp\left(\frac{\lambda_{3,3}}{v}\right)\right] ,
\end{align}
where
\begin{align}
\mu=-\int^{\infty}_{0}\ln\left[1-\exp\left(-2z^{2}\right)\right]dz\approx1.6 .
\end{align}
See Appendix \ref{p3app} for the full details.

The critical density is found from the equation
\begin{align}
1=L^{3}P_{3}(L\times L\times L)\Rightarrow v_{c}\approx\frac{\lambda_{3,3}}{\ln\ln L} .
\end{align}

\subsubsection*{Strong confinement in 3D rhomboids}

In order to estimate the critical density in quasi-planes we consider the typical size of the unfrozen clusters. We consider only square clusters and assume that averaging over all possible shapes is the same as averaging over squares. Similarly to the $m=2$ case in 2D, we assume that the relative probability of clusters of size $l^2$ is the probability to expand a critical droplet to size $L_{1}\times l\times l$. We now fix $L_{1}$ and take $L_{2}$ and $L_{3}$ to infinity, and expand a droplet. Up to size $L_{1}\times L_{1}\times L_{1}$, the expansion is the same as in a bulk system. After that, the droplet can be expanded in only two directions, and each side of the rhomboid is of size $L_{1}\times l$. The probability that each side is unblocked is $\beta^{l}_{3}(L_{1})$, where $\beta_{3}$ was defined in Eq. (\ref{b3}). However, the side of the rhomboid may be divided into subsections, similarly to the division of subsections in the $m=2$ model in two-dimensions, by walls consisting of two or more adjacent full walls of size $L_{1}\times2$. The average length of these subsections is $\approx\rho^{-4L_{1}}$, and above that length some of the sites in the rhomboid's side are very likely to be frozen. Therefore, the characteristic size of the sections is $L_{1}\times \rho^{-4L_{1}}\times \rho^{-4L_{1}}$. Note that this characteristic size does not depend on $\lambda_{3,3}$. The critical density is found from the equation
\begin{align}
v^{QP}_{c}\approx\frac{\lambda_{3,3}}{\ln\ln L_{1}\rho^{-8L_{1}}_{c}}\approx\frac{\lambda_{3,3}}{\ln\left(\ln L_{1}+8L_{1}v_{c}\right)} .
\end{align}
In the limit of $8L_{1}v_{c}\gg\ln L_{1}$, the solution is
\begin{align}
v^{QP}_{c}=\frac{1}{W_{0}\left(8\lambda_{3,3}L_{1}\right)}\approx\frac{1}{\ln\left(8\lambda_{3,3}L_{1}\right)} ,\label{vqpc}
\end{align}
where $W_{0}\left(x\right)$ is the product-log function \cite{productlog}.

When the system is confined to a tunnel, the side of the rhomboid is not divided into subsections, because it is of constant size $L_{1}\times L_{2}$. In that case, the tunnel is divided into sections by walls, which are one or more adjacent planes of size $L_{1}\times L_{2}$ that are frozen. Assuming that $L_{1}$ and $L_{2}$ are large enough, the probability that a plane is frozen is $1-\exp\left[-L_{1}L_{2}\exp\left(-2\lambda_{3,3}/v\right)\right]$. This is very similar to a tunnel in the $m=2$ model in which the probability that a plane is frozen is $1-\rho^{L_{1}L_{2}}\approx1-\exp\left(-vL_{1}L_{2}\right)$. Therefore, we can use the known results for tunnels in $m=2$ and replace $v$ with $\exp\left(-\lambda_{3,3}/v\right)$. The average section length is
\begin{align}
\left\langle l\right\rangle\approx\exp\left[L_{1}L_{2}\exp\left(-2\lambda_{3,3}/v\right)\right] .
\end{align}
The critical density is found from the equation
\begin{align}
&v^{tunnel}_{c}\approx\frac{\lambda_{3,3}}{\ln\ln\left(L_{1}L_{2}\left\langle l\right\rangle\right)} ,
\end{align}
and thus
\begin{align}
&v^{tunnel}_{c}\approx\frac{3\lambda_{3,3}}{\ln\left(L_{1}L_{2}\right)} .\label{vtunnelc}
\end{align}

\section{Jamming by shape in 3D rhomboids in the $m=2$ model}
\label{SectionJamShape}
In this Section we show how changing only the shape of the system can change its behavior and induce jamming. Although we show this only for three-dimensional systems in the $m=2$ model, the conclusions are valid for any dimension and any $m$.

We focus on four possible relations between the three length scales $L_{1},L_{2}$ and $L_{3}$. If $L_{1}\approx L_{2}\approx L_{3}$ the system behaves as a cubic (bulk) system. If $L_{1}\ll L_{2}\approx L_{3}$ we call the system a quasi-planar (QP) system. If $L_{1}\approx L_{2}\ll L_{3}$ the system is a tunnel. If $L_{1}\ll L_{2}\ll L_{3}$ we call the system a quasi-tunnel (QT). The critical vacancy densities in the four regimes are given by (see Section \ref{critm2})
\begin{align}
&v^{bulk}_{c}=\left(\frac{3\lambda_{3,2}}{\ln V}\right)^{2}&v^{QP}_{c}=\frac{2\lambda_{2,2}}{L_{1}\ln V}\nonumber\\
&v^{tunnel}_{c}=\left(\frac{3\lambda_{3,2}}{2L_{1}L_{2}}\right)^{2/3}&v^{QT}_{c}=\sqrt{\frac{\lambda_{2,2}}{L^{2}_{1}L_{2}}} .\label{vcm2d3}
\end{align}

Bulk $(d_{eff}=3)$ and QP $(d_{eff}=2)$ systems are weakly confined, while tunnels $(d_{eff}=1,t=0)$ and QT $(d_{eff}=1,t=1)$ systems are strongly confined. As noted before, a strongly confined system is divided into clusters. In the $m=2$ model, we call these clusters \textit{sections}, as they behave somewhat differently than clusters in the $m>2$ models. Only in the $m=2$ model, the sections are independent of each other, in the manner that the state of a section (whether it is frozen or not) is completely independent of the state of the neighboring sections. These sections are separated by at least two adjacent fully occupied planes perpendicular to the long direction.

By changing the different length scales, $L_{1},L_{2},L_{3}$, for fixed volume $V=L_{1}L_{2}L_{3}$, the system crosses over between the different regimes. For example, by changing the value of $L_{1}$, the system will crossover from bulk behavior to quasi-planar behavior. This transition occurs when the critical density satisfies both the bulk equation and the QP equation. Equating the two yields the crossover length for $L_{1}$
\begin{align}
L^{bulk-QP}_{1}=\frac{2\lambda_{2,2}}{9\lambda^{2}_{3,2}}\ln V .
\end{align}
Similarly, by decreasing $L_{3}$, the system will change from a tunnel into a bulk system or from a quasi-tunnel into a quasi-plane. By equating the relevant expressions for the critical density we find that the crossover length is
\begin{align}
L^{bulk-tunnel}_{3}=\frac{\exp\left[\left(18\lambda^{2}_{3,2}L_{1}L_{2}\right)^{1/3}\right]}{L_{1}L_{2}} .
\end{align}
In Section \ref{critm2}, we showed that the average section size is equal to $L^{bulk-tunnel}_{3}$. Hence, the transition between bulk and tunnel occurs at the point where the system contains on average one section. 
The third crossover length is between tunnels and quasi-tunnels,
\begin{align}
L^{QT-tunnel}_{1}=\left(\frac{2\lambda^{3/2}_{2,2}}{3\lambda_{3,2}}\right)^{2}\sqrt{L_{2}} .
\end{align}

\begin{figure}
\includegraphics[width=200pt]{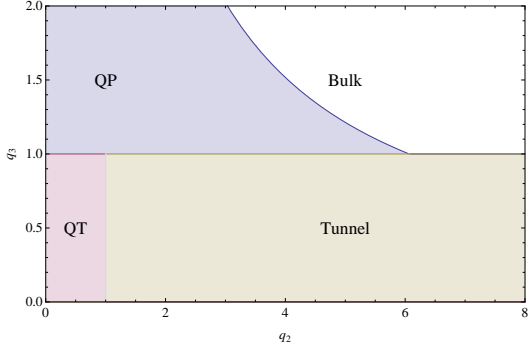}
\caption{Phase diagram of the four possible states of a 3D system (bulk, quasi-plane (QP), tunnel, and quasi-tunnel (QT)), as a function of the two shape parameters $q_{2}=\left(\frac{9\lambda^{2}_{3,2}}{4\lambda^{3}_{2,2}}\right)^{2}\frac{L^{2}_{1}}{L_{2}}$ and $q_{3}=18\lambda^{2}_{3,2}\frac{L_{1}L_{2}}{\ln^{3}V}$.}
\label{phasesm2d3}
\end{figure}

To show this behavior graphically, we define the three following parameters describing the shape of the system
\begin{align}
&q_{1}=\left(\frac{9\lambda^{2}_{3,2}}{2\lambda_{2,2}}\right)^{3}\frac{L^{3}_{1}}{\ln^{3}V} ,\nonumber\\
&q_{2}=\left(\frac{9\lambda^{2}_{3,2}}{4\lambda^{3}_{2,2}}\right)^{2}\frac{L^{2}_{1}}{L_{2}} ,\nonumber\\
&q_{3}=18\lambda^{2}_{3,2}\frac{L_{1}L_{2}}{\ln^{3}V} ,
\end{align}
such that $q_{1}>1$ implies $L_{1}>L^{bulk-QP}_{1}$, $q_{2}>1$ implies $L_{1}>L^{QT-tunnel}_{1}$, and $q_{3}>1$ implies $L_{3}<L^{bulk-tunnel}_{3}$. Also, we note that $q_{2}q_{3}=q_{1}/\lambda^{3}_{2,2}$. Therefore, the state of the system can be determined by any two of the three parameters, $q_{1}, q_{2}$ and $q_{3}$, as shown in Fig. \ref{phasesm2d3}.

As each regime has a different critical density, by changing the system's shape, but not its volume or density, a system may become jammed if the density was lower than the critical density at the original shape, but higher than the critical density at the new shape.

\begin{figure}
\includegraphics[width=\columnwidth]{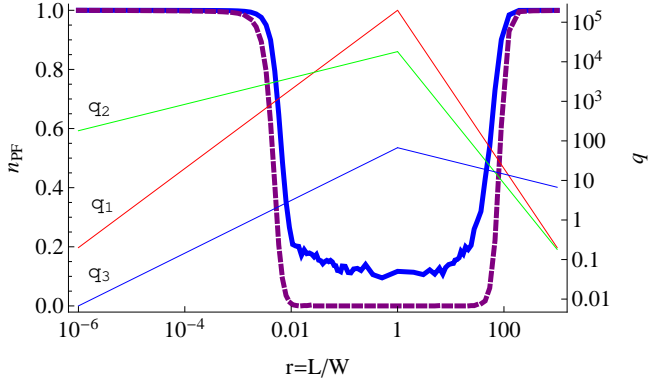}
\caption{Average fraction of frozen particles $n_{PF}$ and the three $q$ parameters vs. aspect ratio $r=L/W$ for system of constant volume $V=WL^{2}=10^6$ and particle density $\rho=0.99$. The thick lines are $n_{PF}$ for the KA (blue solid line) and FA (purple dashed line) models, while the thin lines are the $q_{1}$ (red), $q_{2}$ (green), and $q_{3}$ (blue). The system changes from tunnel to bulk at $r\approx0.01$ ($W\approx2100$, $L\approx21$, $q_{3}=1$) and from bulk to quasi-planar at $r\approx70$ ($W\approx6$, $L\approx420$, $q_{1}\approx10$). The value of $q_{1}$ at the transition from bulk to quasi-planar is not $1$ but larger, because the values of the $\lambda$ parameters are far from their asymptotic value.} 
\label{m2difr}
\end{figure}

Consider a system of size $W\times L\times L$ and define the aspect ratio $r=L/W$. The system has constant volume $V=WL^{2}$ and constant density, $\rho$. By changing the aspect ratio between $0$ and $\infty$ the system changes from a tunnel ($r\ll1$) to a bulk system ($r\approx1$) and to a quasi-plane ($r\gg1$). These transitions can be seen in Fig. \ref{m2difr}, where the aspect ratio is changed in a system of volume $V=10^6$ and density $\rho=0.99$. This density was chosen so that the bulk system is unjammed, but the tunnel and quasi-planar systems are jammed. In the KA model in the bulk, the chosen density is very close to the critical density $\rho\approx\rho_{c}$ and so $n_{PF}\approx0.2$ and not $0$ or $1$. Even near $\rho_{c}$, $n_{PF}$ does not change much with $r$ and depends only on $V$ as long as the system is in the bulk regime. Fluctuations in $n_{PF}$ appear since $V$ is not exactly the same for all aspect ratios, but varies from $V=994194$ at $L=171$ to $V=1011240$ at $L=159$, since the system must be a rhomboid.

\begin{figure}
\includegraphics[width=240pt]{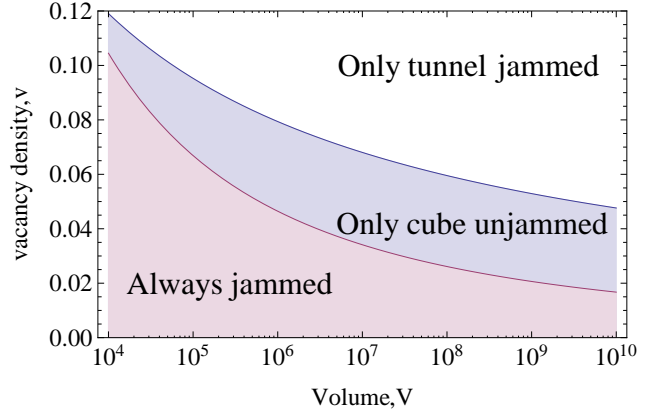}
\caption{The behavior of a 3D system in the $m=2$ models as a function of its volume and vacancy density. This plot was drawn using the asymptotic values of $\lambda_{2,2}$ and $\lambda_{3,2}$. If we knew their values for any volume and density, the plot would be more accurate. Hence, the values chosen in Fig. \ref{m2difr} ($V=10^6,v=0.01$) appear here in the ''Always jammed" region, while they should be in the ''Only cube unjammed" region.}
\label{phases}
\end{figure}

In general, for a given volume $V$ and density $\rho$ there are four possible behaviors: 1) neither a tunnel, a 2D plane nor a cube are jammed; 2) a tunnel is jammed, but not a plane or a cube; 3) a tunnel and a plane are jammed, but not a cube; 4) all possible shapes are jammed. The first behavior, that a tunnel is unjammed, occurs only for very low densities, since the critical density of an infinite tunnel is approximately $0.3$ \cite{teomy}, and the critical densities for large planes and cubes are much higher. Therefore, we disregard that possibility. By varying either the system's volume or the particle density, the behavior of the system changes between the other three possibilities. The crossover occurs along the lines in which the density is equal to the critical density for 2D planes of that volume and for cubes of that volume. Note that for each combination of volume and density, the transition between bulk and quasi-planes occurs at a different aspect ratio. Figure \ref{phases} shows the behavior of the system as a function of its volume and vacancy density. 

\begin{figure}
\includegraphics[width=\columnwidth]{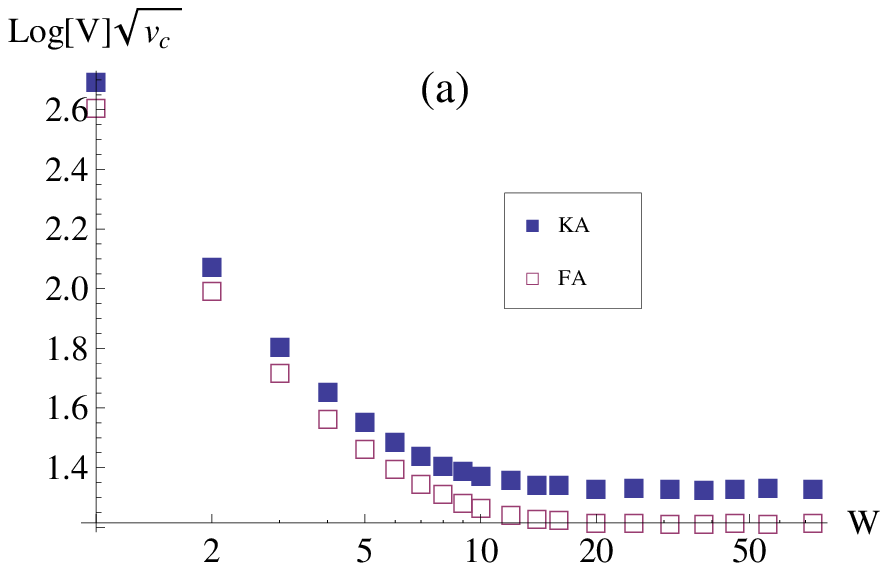}\\
\includegraphics[width=\columnwidth]{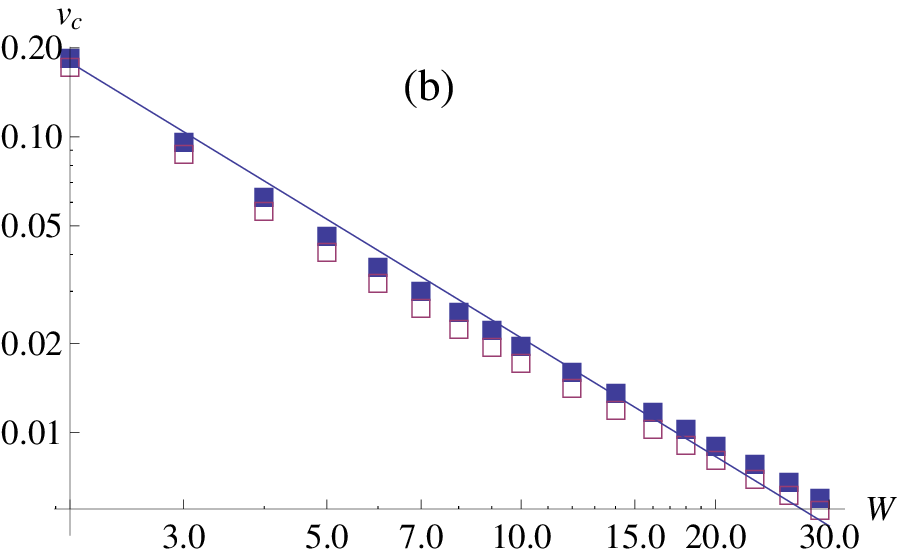}
\caption{(a) The value of $3\lambda_{3,2}=\sqrt{v_{c}}\ln V$ as a function of $W$ for systems of size $W\times10^{3}\times10^{3}$ for the KA and FA models. Above $W\geq15$, $\lambda_{3,2}$ does not depends on $W$, which fits the behavior of a bulk system. Below $W\leq15$, the system is a quasi-plane so it does not correspond to the bulk behavior. (b) The critical vacancy density $v_{c}$ for tunnels of size $W\times W\times L$ as a function of the width $W$ for the KA and FA models. The length $L$ is long enough for the tunnel to be considered infinite ($L=10^3$ for $W\leq9$, $L=2\times10^3$ for $W\leq16$, $L=4\times10^3$ for $W=18$, $L=2\times10^4$ for $W=20,23$, $L=5\times10^4$ for $W=26$, and $L=2\times10^5$ for $W=29$). The straight line is $\sim W^{-4/3}$, which is the expected behavior of tunnels, see Eq. (\ref{vcm2d3})}
\label{m2qp}
\end{figure}

The crossover from QP behavior to bulk behavior is illustrated in Fig. \ref{m2qp} (a), which shows $\sqrt{v_{c}}\ln V$ as a function of $W$ for systems of size $W\times10^3\times10^3$. When $W\geq15$, the value of $\sqrt{v_{c}}\ln V$ is approximately constant, which fits the behavior of bulk systems; see Eq. (\ref{vcm2d3}). In QP systems, we see from Eq. (\ref{vcm2d3}) that $\sqrt{v_{c}}\ln V\sim\sqrt{\ln (V)/W}$ which qualitatively agrees with the results for $W\leq15$. However, for small values of $W$ the simulation results do not scale as expected, because in these relatively small systems the effective value of $\lambda$ depends strongly on the system's size.

Figure \ref{m2qp} (b) shows the critical vacancy density of tunnels of size $W\times W\times L$, where $L$ is large enough so that the system is practically infinitely long. We see that even for $W$ as small as $2$, the critical vacancy density scales as $W^{-4/3}$, as expected from Eq. (\ref{vcm2d3}). 

\section{Correlation length in strongly confined 3D rhomboids in the $m=3$ model}
\label{SectionCorrelation}
\begin{figure}
\subfigure{\includegraphics[width=197pt]{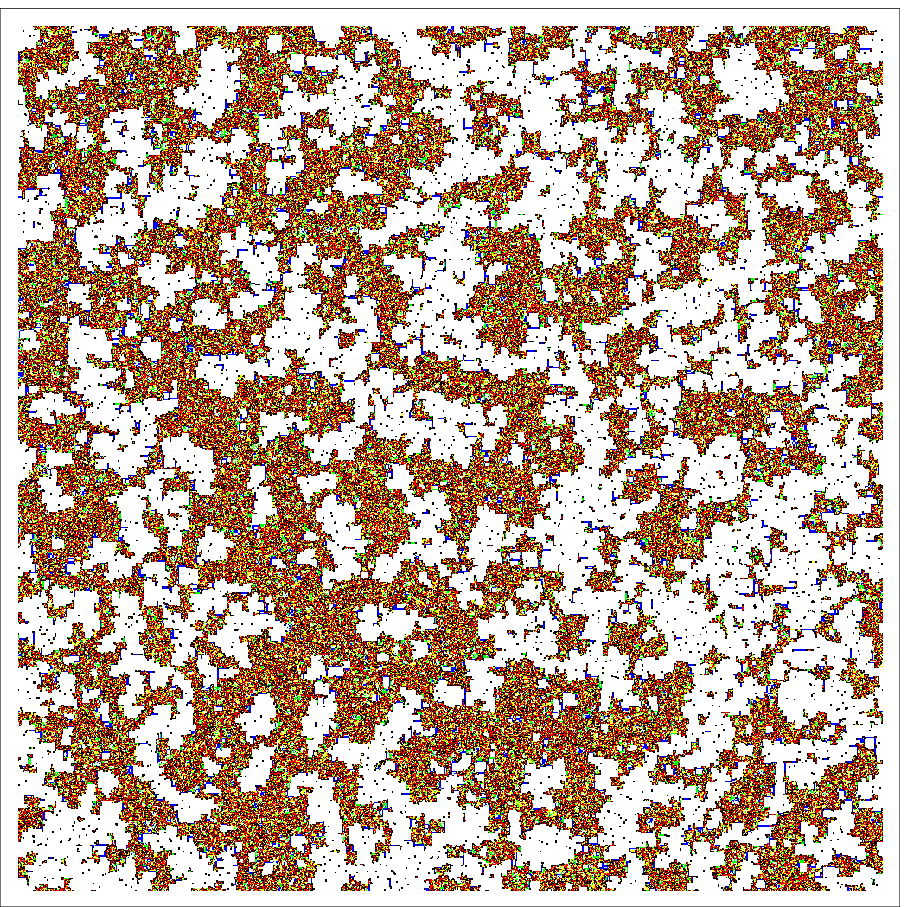}}\\
\subfigure{\includegraphics[width=197pt]{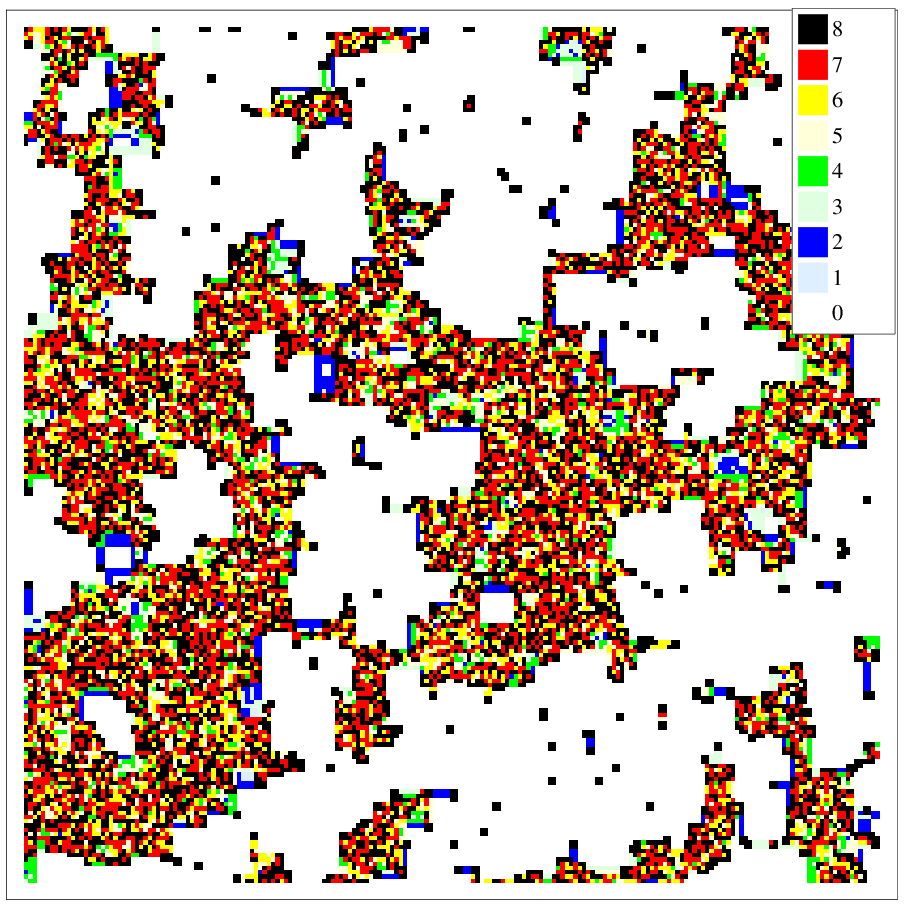}}
\caption{Two-dimensional projection of the frozen particles in a system of size $1000\times1000\times8$ in the FA model with $m=3$ with periodic boundary conditions. The density is $\rho=0.881$ and the fraction of frozen particles is $n_{PF}=0.46$. The colors represent the number of frozen particles in each column, as shown in the lower panel. The full columns do not create closed sections, but when they are close they create a frozen cluster by scaffolding. The second panel is a zoom-in on a portion $200\times200$ of the projection.}
\label{3dcon}
\end{figure}

In the $m=3$ model, a strongly confined system is divided into frozen clusters which are not independent. Consider a ``view from above'' on a three-dimensional system, such that it appears to be a two-dimensional system, say in the $x-y$ plane. Each ``site'' in this effective 2D system represents a column which is either completely full or not completely full. Any closed shape formed by full columns is frozen, and the sites confined by this wall are independent of the sites outside the wall. The most probable closed shape is a $2\times2$ square. If two such vertical pillars are close to each other, there is a possibility that they are connected by a horizontal scaffold, which is also frozen if it is at least two sites wide. These scaffolds can also connect between themselves, and thus form frozen clusters. The clusters are not completely independent, but their correlation decreases with the distance between them. Figure \ref{3dcon} shows an example of a configuration and the frozen sections in it. The 2D projection of the frozen clusters is very similar to the results of continuous models in 2D \cite{einav}, in which particles form dense regions reminiscent of our frozen clusters. Moreover, with hard-wall boundary conditions both the frozen clusters in the KA and FA models and the dense regions in the continuous model form near the edges of the system.

\begin{figure}
\subfigure{\includegraphics[width=\columnwidth]{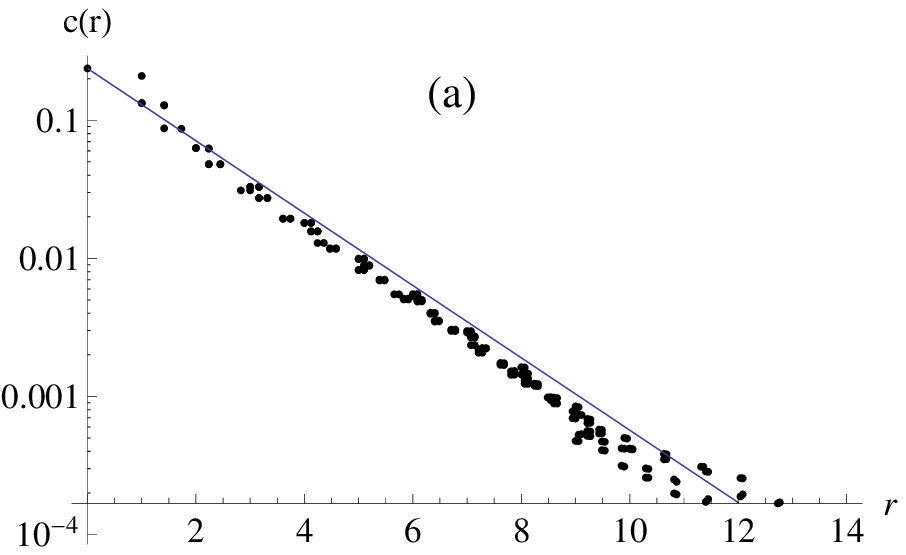}}\\
\subfigure{\includegraphics[width=\columnwidth]{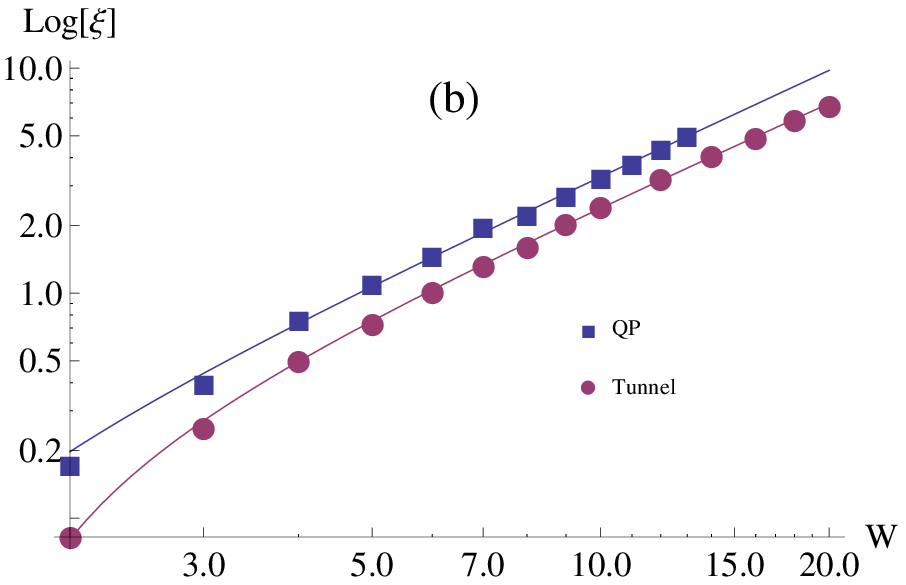}}
\caption{(a) The correlation $c_{\vec{i},\vec{j}}$ as a function of the distance $r=\left|\vec{i}-\vec{j}\right|$ for a quasi-plane of size $3\times100\times100$ in the FA model near the critical density $\rho=0.81$. The straight line is $\rho n_{PF}\left(1-\rho n_{PF}\right)\exp\left[-r/\xi\right]$, with $\xi\approx1.66$ calculated by the fluctuations in $n_{PF}$. The deviation from the line is due to the large weight of the few nearby points whose correlation requires corrections to the exponential decay. (b) 
The estimated correlation length $\xi$ as a function of the width $W$ for quasi-planes of size $W\times\infty\times\infty$ (squares) and tunnels of size $W\times W\times\infty$ (circles) at the critical density. The continuous lines are a fit to $\xi=\gamma_{1}\exp\left[\gamma_{2}W^{\gamma_{3}}\right]$.}
\label{corvdis}
\end{figure}

As an approximation, we consider the frozen areas as adjacent frozen clusters of typical size $\xi$, and similarly for the unfrozen areas. We assume that at the critical density, the typical size of the frozen and unfrozen clusters is the same. We further assume that the typical cluster size is equal to the correlation length between frozen sites. In order to obtain the correlation length, we assume that the correlation between sites $\vec{i}$ and $\vec{j}$ being frozen decreases exponentially with their distance
\begin{align}
&c_{\vec{i},\vec{j}}=\rho n_{PF}\left(1-\rho n_{PF}\right)\exp\left[-\frac{\left|\vec{i}-\vec{j}\right|}{\xi}\right] .\label{cor}
\end{align}
The numerical results, shown in Fig. \ref{corvdis} (a), support this assumption. Finding the correlation length numerically using Eq. (\ref{cor}) is very costly. Instead, we find it using a related quantity which is easier to evaluate numerically, namely the average correlation in the system
\begin{align}
&C_{av}=\frac{1}{V^{2}}\sum_{\vec{i},\vec{j}}c_{\vec{i},\vec{j}}=\frac{c_{0}}{V}\sum_{x,y,z}\exp\left[-\sqrt{x^{2}+y^{2}+z^{2}}/\xi\right] ,\label{eqcav}
\end{align}
where $c_{0}=\rho n_{PF}\left(1-\rho n_{PF}\right)$, $x,y,z$ are the distance along the three directions between two sites, and when using periodic boundary conditions, the three sums are from $-\left\lfloor (L_{i}+1)/2\right\rfloor+1$ to $\left\lfloor L_{i}/2\right\rfloor$, with $\left\lfloor q\right\rfloor$ being the integer part of $q$.  

Appendix \ref{snpf} shows that $C_{av}$ is related to the fluctuations in the fraction of frozen particles $n_{PF}$ between different configurations by
\begin{align}
C_{av}=\sigma^{2}=\frac{1}{A}\sum_{\alpha}\left(\rho n^{\alpha}_{PF}\right)^{2}-\left(\frac{1}{A}\sum_{\alpha}\rho n^{\alpha}_{PF}\right)^{2} ,\label{sig2}
\end{align}
where the summation is over $A$ different realizations $\alpha$ of the system, and $n^{\alpha}_{PF}$ is the fraction of frozen particles in configuration $\alpha$.

In quasi-planes, we assume that $L_{2},L_{3}\gg\xi$, such that the sums of $y$ and $z$ in Eq. (\ref{eqcav}) may be changed to integrals with the limits of the integrals taken to infinity. We now change the integration variables from $y$ and $z$ to $r=\sqrt{y^{2}+z^{2}}$ and $\theta=\tan^{-1}z/y$, such that
\begin{align}
&C^{QP}_{av}=\frac{c_{0}}{V}\sum_{x}\int^{\infty}_{0}\int^{2\pi}_{0}rdrd\theta\exp\left[-\sqrt{x^{2}+r^{2}}/\xi\right]=\nonumber\\
&=\frac{2\pi c_{0}}{V}\sum^{\left\lfloor L_{1}/2\right\rfloor}_{x=-\left\lfloor(L_{1}+1)/2\right\rfloor+1}\xi\left(\xi+|x|\right)\exp\left[-|x|/\xi\right]=\nonumber\\
&=\frac{\pi\xi c_{0}}{V\sinh^{2}\left(1/2\xi\right)}\left\{1+\xi\sinh\left(1/\xi\right)-\right.\nonumber\\
&-\left.\exp\left[-\frac{\left\lfloor L_{1}/2\right\rfloor}{\xi}\right]\left[1+\left(\xi+\left\lfloor L_{1}/2\right\rfloor\right)\times\right.\right.\nonumber\\
&\left.\left.\times\left(\sinh\left(1/\xi\right)-2\delta_{L_{1},odd}\sinh^{2}\left(1/2\xi\right)\right)\right]\right\} .\label{cqpav}
\end{align}
For wide quasi-planes we find that $\xi\gg L_{1}$, such that
\begin{align}
C^{QP}_{av}\approx\frac{2\pi\xi^{2}L_{1}c_{0}}{V} .
\end{align}

In tunnels we assume that $L_{3}\gg\xi$, such that the sum of $z$ in Eq. (\ref{eqcav}) may be changed to an integral with the limits of the integral taken to infinity. Hence
\begin{align}
&C^{tun}_{av}=\frac{c_{0}}{V}\sum_{x,y}\int^{\infty}_{-\infty}\exp\left[-\sqrt{x^2+y^2+z^2}/\xi\right]dz=\nonumber\\
&=\frac{2c_{0}}{V}\sum_{x,y}\int^{\infty}_{0}\exp\left[-\sqrt{x^2+y^2+z^2}/\xi\right]dz .
\end{align}
changing the integration variable to $r=\sqrt{x^{2}+y^{2}+z^{2}}$ yields
\begin{align}
&C^{tun}_{av}=\frac{2c_{0}}{V}\sum_{x,y}\int^{\infty}_{\sqrt{x^{2}+y^{2}}}\exp\left[-r/\xi\right]\frac{rdr}{\sqrt{r^{2}-x^{2}-y^{2}}}=\nonumber\\
&=\frac{2c_{0}}{V}\sum_{x,y}\sqrt{x^{2}+y^{2}}K_{1}\left(\frac{\sqrt{x^{2}+y^{2}}}{\xi}\right) ,\label{ctunav}
\end{align}
where $K_{1}$ is the modified Bessel function of the second kind \cite{bessel}. The sums over $x$ and $y$ are done numerically. In wide tunnels, we also find that $\xi\gg L_{1},L_{2}$, such that $rK_{1}(r/\xi)\approx\xi$, and thus
\begin{align}
&C^{tun}_{av}\approx\frac{2c_{0}\xi}{V}\sum_{x,y}1=\frac{2\xi L_{1}L_{2}c_{0}}{V} .
\end{align}

By calculating $\sigma^{2}$ numerically, and using Eq. (\ref{sig2}) and either Eq. (\ref{cqpav}) (for quasi-planes) or (\ref{ctunav}) (for tunnels), we found the correlation length, $\xi$.
We fitted the value of $\xi$ to a function of the form
\begin{align}
\xi=\gamma_{1}\exp\left[\gamma_{2}W^{\gamma_{3}}\right] ,\label{eqxi}
\end{align}
where $W$ is the width of the system for either a quasi-plane of size $W\times\infty\times\infty$ or a tunnel of size $W\times W\times\infty$. The result is shown in Fig. \ref{corvdis} (b). As shown in Table \ref{gamtab}, the values of the $\gamma$ parameters for quasi-planes and tunnels are similar. Note that the correlation length in Eq. (\ref{eqxi}) has a similar functional form as the average section length in tunnels in the $m=2$ model (see Section \ref{critm2})
\begin{align}
\left\langle l\right\rangle=\exp\left[2^{-1/d}\left(d\lambda_{d,2}\right)^{1-1/d}W^{1-1/d}\right] ,
\end{align}
such that for $d=3$, $\gamma_{1}=1$, $\gamma_{2}\approx1.64$, and $\gamma_{3}=2/3$.

\begin{table}
\begin{tabular}{l|c|c|c}
&$\gamma_{1}$&$\gamma_{2}$&$\gamma_{3}$\\\hline
Quasi-plane&$0.923$&$0.0948$&$1.55$\\
Tunnel&$0.861$&$0.0861$&$1.49$
\end{tabular}
\caption{The values of the $\gamma$ parameters obtained by the numerical fit to $\xi=\gamma_{1}\exp\left[\gamma_{2}W^{\gamma_{3}}\right]$ at the critical density.}
\label{gamtab}
\end{table}

As an approximation, we assume that a cluster of size $\xi$ is unfrozen if it contains a critical droplet, such that the critical density may be obtained for large $W$ from the relation
\begin{align}
1=\xi^{2}W P\left[\xi\times \xi\times W\right] 
\end{align}
for quasi-planes, and
\begin{align}
1=\xi W^{2}P\left[\xi\times W\times W\right]
\end{align}
for tunnels. Solving these equations for large $W$ yields in both cases
\begin{align}
v_{c}=\frac{\lambda_{3,3}}{\gamma_{3}\ln W} ,\label{vcqpm3}
\end{align}
which is of the same functional form as the critical density for two dimensional systems in the $m=2$ model but with a different prefactor, and agrees qualitatively with Eqs. (\ref{vqpc}) and (\ref{vtunnelc}). Interestingly, we obtain the same expression for quasi-planes and for tunnels. We cannot compare Eq. (\ref{vcqpm3}), which is valid for very large systems, to our numerical results, because our simulations are done on relatively small systems ($W=13$ for QP and $W=20$ for tunnels) in which the effective value of $\lambda_{3,3}$ varies significantly and is very far from its asymptotic value.

\subsection*{Transition from bulk to quasi-planar}

\begin{figure}
\includegraphics[width=0.45\columnwidth]{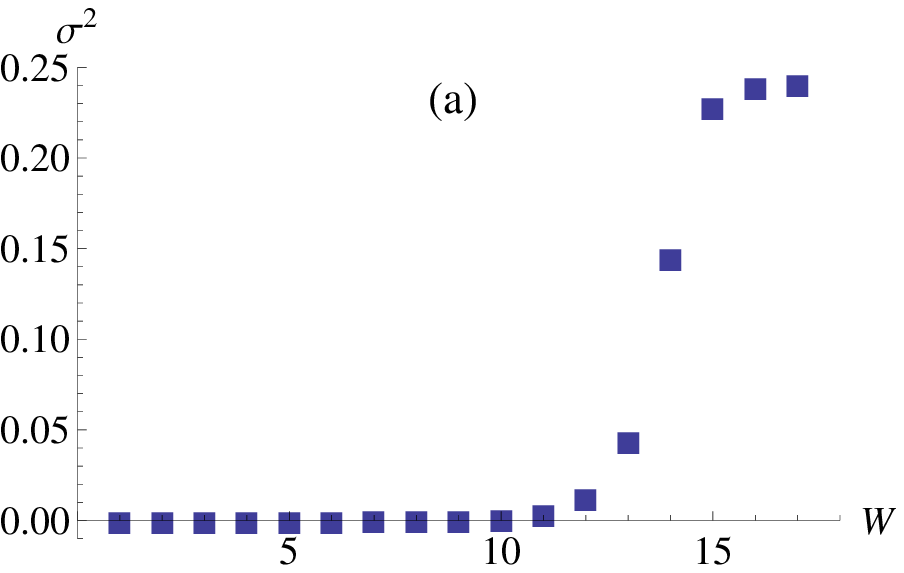}
\includegraphics[width=0.45\columnwidth]{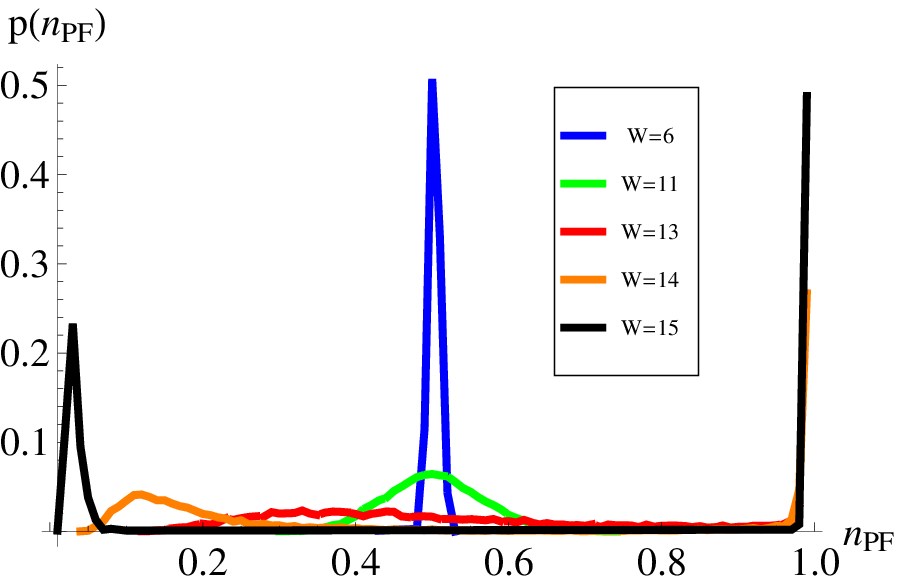}
\caption{(a) The variance $\sigma^{2}$ in the fraction of frozen particles and (b) the distribution $p(n_{PF})$ at the critical density for a system of size $W\times1000\times1000$ in the $m=3$ FA model. At small width ($W=6, \rho_{c}=0.8536$, blue curve in panel (b)) there is a sharp peak around $0.5$. Near the crossover region ($W=11, \rho_{c}=0.88806$, green curve) the peak is still visible, but is much broader. During the crossover ($W=13, \rho_{c}=0.89661$, red curve) the distribution almost flattens, but the emergence of the peak at $1$ can be seen. At $W=14$ ($\rho_{c}=0.90006$, orange curve) the peak at $1$ is clearly seen, and a broad peak centered around $0.1$ is also visible. As the width is further increased ($W=15, \rho_{c}=0.90205$, black curve) the system becomes almost bulk and the two peaks at $1$ and near $0$ are obvious. At that point $(W=15)$ there is a crossover from quasi-planar behavior, in which $\sigma^{2}\approx0$, to a bulk behavior, in which $\sigma^{2}\approx1/4$.}
\label{m3qpsig}
\end{figure}

Since the critical density changes very slightly in the system sizes we investigated, another technique was used to see the crossover between bulk behavior and quasi-planar behavior. In the bulk, the system is either almost completely unfrozen or almost completely frozen, and thus its probability of being frozen can be approximated by a binary distribution. In this case, the variance, $\sigma^{2}$, of the fraction of frozen particles over many configurations is approximately $\sigma^{2}\approx n_{PF}(1-n_{PF})$, or $1/4$ at the critical density. In a two-dimensional system, however, the frozen structures are very local, and thus a large enough system can in itself be considered an average over many small systems. In this case, the distribution of $n_{PF}$ over the different configurations is almost constant and thus the variance is almost zero. The distribution of $n_{PF}$ and its variance at the critical density are plotted in Fig. \ref{m3qpsig} for a system of size $W\times1000\times1000$. At that size, the crossover occurs at $W\approx15$, which means that for systems of width $15$, the characteristic size of the sections is of the same order of magnitude as the system size. The characteristic cluster size, calculated by Eq. (\ref{eqxi}) is appproximately $\xi\approx500$, which is expected given the system size $(L=1000)$. For small widths $(W\leq11)$, there is a pronounced peak in the distribution at $0.5$, which becomes broader as the width increases. At larger widths, the emergence of the two peaks at $0$ and $1$ is visible. Note also that the distribution is not symmetric: while the peak at $1$ is very large, the peak near $0$ is broad and is centered around a small positive value, which nears $0$ as the width increases. This means that the number of frozen particles in the almost unfrozen configurations vary, while those that are almost completely frozen have very small and negligible unfrozen regions.

\section{Summary}
\label{SectionSummary}

In this paper we investigated the effects of the system's shape on the jamming transition. We derived an analytical approximation for the critical density in a $d$-dimensional hyper-rhomboid system in both the Kob-Andersen and Fredrickson-Andersen kinetically-constrained models, and showed that it scales differently with the system's length scales depending on the relation between them. We distinguished between two general classes of systems depending on the model's parameter $m$ and the number of effective dimensions in the systems, $d_{eff}$: a weakly confined system $(d_{eff}\geq m)$ and a strongly confined system $(d_{eff}<m)$. In the weakly confined regime, the system is either completely unfrozen or almost completely frozen, and the correlation length between frozen sites is the entire system size. The critical density in a weakly confined system depends on the volume of the system $V$ and the volume associated with the $s$ small dimensions $V_{s}$, but not on the ratios between the small dimensions or the large dimensions. In the strongly confined regime, the system is divided into frozen and unfrozen clusters, such that the correlation length between frozen sites is the size of the clusters, not of the system. The critical density in a strongly confined system depends on $V_{s}$, but not on the large dimensions or the ratios between the small dimensions.

We also showed how changing the system's shape without altering its total volume or the particle density can induce jamming. This was done by utilizing our result that the critical density depends on the system's shape, such that a certain density may be below the critical density at a particular shape but above it for a different shape. We emphasize that this is derived by averaging over ensembles of different realizations, not by exerting forces on the system. Although we considered only the Kob-Andersen and the Fredrickson-Andersen models, this conclusion may be applicable to other models as well, including continuous models. 

The results presented in this paper regarding the effect of the shape on the static properties of the models will be the groundwork for an investigation of the effect of the shape on the dynamic properties of these models.

\section*{Acknowledgements}

We thank Roman Golkov and Itai Einav for helpful discussions. This research was supported by the Israel Science Foundation grants No. $617/12$, $1730/12$.

\appendix

\section{Proof that sections in the $m=2$ models in $d=3$ cannot be further divided}
\label{nodivisions}

In order for a section to be divided into subsections, it needs to satisfy the condition 
\begin{align}
L_{2}>\exp\left[2vL_{1}\left\langle l_{3}\right\rangle\right] ,
\end{align}
or equivalently
\begin{align}
vL_{1}<\frac{W_{0}\left(L_{2}\ln L_{2}\right)}{2L_{2}} ,
\end{align}
where $W_{0}(z)$ is the product-log function \cite{productlog} defined as the solution to
\begin{align}
z=W_{0}(z)\exp\left[W_{0}(z)\right] .
\end{align}
For large $L_{2}$ we can approximate this by
\begin{align}
vL_{1}<\frac{\ln L_{2}}{2L_{2}} .
\end{align}
The critical vacancy density in this case is larger than or equal to that of a tunnel (because the subsection is smaller and thus easier to jam), and so the condition can be written as
\begin{align}
\left(\frac{3\lambda_{3,2}}{2L_{1}L_{2}}\right)^{2/3}L_{1}<vL_{1}<\frac{\ln L_{2}}{2L_{2}} ,
\end{align}
or equivalently
\begin{align}
L_{1}<\frac{\ln^{3}L_{2}}{9\lambda^{2}_{3,2}L_{2}} .
\end{align}
The right hand side is bounded by $3/\left(e^{3}\lambda^{2}_{3,2}\right)\approx0.15$. This means that $L_{1}$ cannot possibly satisfy the condition, and therefore no further partitions are possible.

\section{Calculation of $P_{3}$}
\label{p3app}
In this section we show that
\begin{align}
&P_{3}(L\times L\times L)\approx\exp\left[-3\mu\exp\left(\frac{\lambda_{3,3}}{v}\right)\right] .
\end{align}

In Eq. (\ref{3dexp}), the contribution to the main sum from small $l$ is negligible, so we can take the upper limit of the inner sum on $k$ to be $\infty$ and approximate it by an integral over $z=kv$, using the approximation $\rho^{k}\approx e^{-vk}$,
\begin{align}
&P_{3}(L\times L\times L)=\nonumber\\
&\exp\left\{3\sum^{L}_{l=1}\ln\left[1-\left(1-\exp\left[-\frac{2}{v}\int^{\infty}_{0}g_{2}(z)dz\right]\right)^{2l^{2}}\right]\right\}\nonumber\\
&=\exp\left\{3\sum^{L}_{l=1}\ln\left[1-\left(1-\exp\left[-\frac{2\lambda_{3,3}}{v}\right]\right)^{2l^{2}}\right]\right\}\approx\nonumber\\
&\approx\exp\left\{3\sum^{L}_{l=1}\ln\left[1-\exp\left(-2l^{2}\exp\left[-\frac{2\lambda_{3,3}}{v}\right]\right)\right]\right\} .
\end{align}
We now change the sum over $l$ to an integral over $z=l\exp\left[-\frac{\lambda_{3,3}}{v}\right]$ and take the upper limit to be $\infty$
\begin{align}
&P_{3}(L\times L\times L)\approx\nonumber\\
&\approx\exp\left\{3\exp\left(\frac{\lambda_{3,3}}{v}\right)\int^{\infty}_{0}\ln\left[1-\exp\left(-2z^2\right)\right]dz\right\}=\nonumber\\
&=\exp\left[-3\mu\exp\left(\frac{\lambda_{3,3}}{v}\right)\right] ,
\end{align}
where
\begin{align}
\mu=-\int^{\infty}_{0}\ln\left[1-\exp\left(-2z^{2}\right)\right]dz\approx1.6 .
\end{align}

\section{Derivation of the Critical Density in Hyper-rhomboids}
\label{critd4}

\subsection{Hyper-rhomboids in the $m=2$ model}
\label{critm2d4}
We consider a hyper-rhombus of size $L_{1}\times...\times L_{d}$, such that $L_{i+1}\geq L_{i}$ for all $i$. The treatment here is very similar to the three-dimensional case. For brevity we define
\begin{align}
V_{k}=\prod^{k}_{i=1}L_{i} .
\end{align}

The expansion of the critical droplet is done analogously to the three-dimensional case such that the probability of the hyper-rhomboid to be unfrozen is
\begin{align}
P=\prod^{d}_{i=1}\prod^{L_{i}}_{l=L_{i-1}+1}\beta^{d-i+1}\left(l^{d-i}V_{i-1}\right) ,\label{phypm2}
\end{align}
where $L_{0}=0$ and $V_{0}=1$.
Setting this in the recursion relation on $P$ yields
\begin{align}
\beta(x)=\frac{1-\rho^{x}+\sqrt{1+2\rho^{x}-3\rho^{2x}}}{2} .
\end{align}
The equation for the critical density is then
\begin{align}
&0=\ln V_{d}+\sum^{d}_{i=1}(d-i+1)\sum^{L_{i}}_{l=L_{i-1}+1}\ln\beta\left(l^{d-i}V_{i-1}\right)\approx\nonumber\\
&\approx\ln V_{d}-\nonumber\\
&-\sum^{d}_{i=1}\frac{d-i+1}{\left(vV_{i-1}\right)^{1/(d-i)}}\int^{\left(vV_{i-1}\right)^{1/(d-i)}L_{i}}_{\left(vV_{i-1}\right)^{1/(d-i)}\left(L_{i-1}+1\right)}g_{1}\left(z^{d-i}\right)dz .\label{eqvgen}
\end{align}

We now check whether the system is divided into sections or not. It is not divided if $L_{d}<\left\langle l_{d}\right\rangle$, for which
\begin{align}
\left\langle l_{d}\right\rangle=\exp\left(2vV_{d-1}\right) .
\end{align}
Assuming that it is not divided, we further assume that there is a $0\leq s<d-1$ such that $\left(vV_{s-1}\right)^{1/(d-s)}L_{s}\ll1$ and $\left(vV_{s}\right)^{1/(d-s-1)}L_{s+1}\gg1$. Then, we can approximate Eq. (\ref{eqvgen}) by
\begin{align}
&0=\ln V_{d}-\frac{d-s}{\left(vV_{s}\right)^{1/(d-s-1)}}\int^{\infty}_{0}g\left(z^{d-s-1}\right)dz=\nonumber\\
&=\ln V_{d}-\frac{(d-s)\lambda_{d-s,2}}{\left(vV_{s}\right)^{1/(d-s-1)}} .
\end{align}
The critical density in this case is
\begin{align}
v_{c}=\frac{1}{V_{s}}\left(\frac{(d-s)\lambda_{d-s,2}}{\ln V_{d}}\right)^{d-s-1} ,
\end{align}
which is similar to the behavior of a hyper-cube in $d_{eff}=d-s$ dimensions with a single length scale, Eq. (\ref{vcgen1}).

If the system is divided into sections, i.e. $L_{d}>\left\langle l_{d}\right\rangle$, it is not further divided into smaller subsections for the exact same reasons as in three-dimensional systems. We can again assume that there is a $0\leq t<d-1$ that satisfies the above conditions, such that the critical density satisfies the equation
\begin{align}
v_{c}=\frac{1}{V_{t}}\left(\frac{(d-t)\lambda_{d-t,2}}{\ln \left(V_{d-1}\left\langle l_{d}\right\rangle\right)}\right)^{d-t-1} .\label{eqvtunnelgen}
\end{align}
If $V_{d-1}$ is large enough such that $\ln V_{d-1}\ll 2v_{c}V_{d-1}$, the solution to Eq. (\ref{eqvtunnelgen}) can be approximated by
\begin{align}
v_{c}=\frac{1}{V_{t}}\left(\frac{(d-t)\lambda_{d-t,2}}{2V_{d-1}}\right)^{1-1/\left(d-t\right)} .
\end{align}

\subsection{Hyper-rhomboids in the $m=3$ model}
\label{critm3d4}

We consider a hyper-rhomboid of arbitrary size and repeat the derivation of the critical droplet. In a hyper-rhomboid, each hyper-side consists of two layers which must be unblocked. Up to size $L_{1}$, the derivation is the same as in the bulk. Above that size and up to size $L_{2}$, there are only $d-1$ directions to consider, and one of the sides of the hyper-side is set to $L_{1}$. Above $L_{2}$, there are only $d-2$ sides, etc. In analogy to the derivation of critical droplets in cubes, Eq. (\ref{3dexp}), and in hyper-rhomboids in the $m=2$ models, Eq. (\ref{phypm2}), the probability that a site is part of a critical droplet is
\begin{widetext}
\begin{align}
&P_{3}=\prod^{L_{1}}_{l=1}\left[1-\left(1-\prod^{l}_{k=1}\beta^{d-1}_{3}(k^{d-2})\right)^{2l^{d-1}}\right]^{d}\prod^{L_{2}}_{l=L_{1}+1}\left[1-\left(1-\prod^{l}_{k=1}\beta^{d-2}_{3}(k^{d-3}L_{1})\right)^{2l^{d-2}L_{1}}\right]^{d-1}\times...\times\nonumber\\
&\times\prod^{L_{d-1}}_{l=L_{d-2}+1}\left[1-\left(1-\prod^{l}_{k=1}\beta_{3}(V_{d-2})\right)^{2lV_{d-2}}\right]^{2}=\nonumber\\
&=\prod^{d-1}_{i=1}\prod^{L_{i}}_{l=L_{i-1}+1}\left[1-\left(1-\prod^{l}_{k=1}\beta^{d-i}_{3}(k^{d-1-i}V_{i-1})\right)^{2l^{d-i}V_{i-1}}\right]^{d+1-i} .
\end{align}
The equation for the critical density is
\begin{align}
&0=\ln V_{d}+\sum^{d-1}_{i=1}\left(d-i+1\right)\sum^{L_{i}}_{l=L_{i-1}+1}\ln\left\{1-\left[1-\exp\left((d-i)\sum^{l}_{k=1}\ln\beta_{3}(k^{d-i-1}V_{i-1})\right)\right]^{2l^{d-i}V_{i-1}}\right\}\approx\nonumber\\
&\approx\ln V_{d}+\sum^{d-1}_{i=1}\left(d-i+1\right)\nonumber\\
&\sum^{L_{i}}_{l=L_{i-1}+1}\ln\left\{1-\left[1-\exp\left(\frac{(i-d)}{\left(vV_{i-1}\right)^{1/(d-i-1)}}\int^{\left(vV_{i-1}\right)^{1/(d-i-1)}l}_{\left(vV_{i-1}\right)^{1/(d-i-1)}}g_{2}\left(z^{d-i-1}\right)\right)\right]^{2l^{d-i}V_{i-1}}\right\} ,\label{eqvgen3}
\end{align}
where we changed the sum over $k$ to an integral over $z=k\left(vV_{i-1}\right)^{1/(d-i-1)}$. We now assume that the system is not strongly confined, and that there are $s$ small sides such that $\left(vV_{s-1}\right)^{1/(d-s-1)}L_{s}\ll1$ and $\left(vV_{s}\right)^{1/(d-s-2)}L_{s+1}\gg1$. The equation for the critical density can then be further approximated by
\begin{align}
&0=\ln V_{d}+\left(d-s\right)\sum^{L_{s+1}}_{l=1}\ln\left\{1-\left[1-\exp\left(\frac{(s+1-d)}{\left(vV_{s}\right)^{1/(d-s-2)}}\int^{\infty}_{0}g_{2}\left(z^{d-s-2}\right)dz\right)\right]^{2l^{d-s-1}V_{s}}\right\}=\nonumber\\
&0=\ln V_{d}+\left(d-s\right)\sum^{L_{s+1}}_{l=1}\ln\left\{1-\left[1-\exp\left(\frac{\lambda_{d-s,3}(s+1-d)}{\left(vV_{s}\right)^{1/(d-s-2)}}\right)\right]^{2l^{d-s-1}V_{s}}\right\}\approx\nonumber\\
&\approx\ln V_{d}+\left(d-s\right)\sum^{L_{s+1}}_{l=1}\ln\left\{1-\exp\left[-2l^{d-s-1}V_{s}\exp\left(\frac{\lambda_{d-s,3}(s+1-d)}{\left(vV_{s}\right)^{1/(d-s-2)}}\right)\right]\right\} .
\end{align}
Changing the sum over $l$ to an integral over $z=lV^{1/(d-s-1)}_{s}\exp\left[-\lambda_{d-s,3}/\left(vV_{s}\right)^{1/(d-s-2)}\right]$ yields
\begin{align}
&0=\ln V_{d}+\frac{\left(d-s\right)\exp\left[\lambda_{d-s,3}/\left(vV_{s}\right)^{1/(d-s-2)}\right]}{V^{1/(d-s-1)}_{s}}\int^{\infty}_{0}\ln\left\{1-\exp\left[-2z^{d-s-1}\right]\right\}dz.
\end{align}
\end{widetext}
Solving this equation yields
\begin{align}
v_{c}\approx\frac{1}{V_{s}}\left(\frac{\lambda_{d-s,3}}{\ln\ln V_{d}}\right)^{d-s-2} .
\end{align}

If the two largest sides, $L_{d}$ and $L_{d-1}$, are much larger than the other sides, i.e. $d_{eff}=2$, the system is quasi-2D. In this case, since the expansion of the droplet in $m=3$ requires at least three dimensions, the droplet can only unblock a finite region. Hence, the system is divided into clusters, each of them either contains a droplet or not. At the critical density we assume that the size of the frozen and unfrozen clusters is the same and equal to $\rho^{-4V_{d-2}}$, such that the critical density satisfies the equation
\begin{align}
&v_{c}\approx\frac{1}{V_{t}}\left(\frac{\lambda_{d-t,3}}{\ln\ln\left(V_{d-2}\exp\left[8v_{c}V_{d-2}\right]\right)}\right)^{d-t-2}\approx\nonumber\\
&\approx\frac{1}{V_{t}}\left(\frac{\lambda_{d-t,3}}{\ln 8v_{c}V_{d-2}}\right)^{d-t-2} .
\end{align}

If there is one large side, $d_{eff}=1$, the system behaves as a tunnel. Similarly to tunnels in three-dimensions, the average section length is
\begin{align}
\left\langle l\right\rangle=\exp\left\{2V_{d-1}\exp\left[-\frac{(d-t-1)\lambda_{d-t,3}}{\left(vV_{t}\right)^{1/(d-t-2)}}\right]\right\} ,
\end{align}
and thus the critical density satisfies the equation
\begin{align}
&v_{c}\approx\frac{1}{V_{t}}\left(\frac{\lambda_{d-t,3}}{\ln\ln\left(V_{d-1}\left\langle l\right\rangle\right)}\right)^{d-t-2}\Rightarrow\nonumber\\
&v_{c}=\frac{1}{V_{t}}\left(\frac{(d-t)\lambda_{d-t,3}}{\ln\left[2V_{d-1}\right]}\right)^{d-t-2} ,
\end{align}
which is similar to a $d-t-1$ dimensional system in the $m=2$ model.

\subsection{Critical density in general hyper-rhomboids}
\label{critm4}
Since the full derivation of the critical density in $m>3$ is very cumbersome, we only provide here a sketch which is very similar to the derivation of the critical density in $m=3$.

Consider a hyper-cube in $d\geq m$ dimensions. This hyper-cube can be expanded if each of its sides has two hyper-layers of $d-1$ dimensions which can be emptied in a manner similar to a $m-1$ model. Now look at the $d-2$ dimensional sides of these layers. They must be emptied in a manner similar to a $m-2$ model, etc. At the end, there is a $d-m+2$ dimensional hyper-surface which must be emptied in a manner similar to a $m=2$ model. Each such iteration of surfaces of lower dimensions adds another exponent to the probability of a site to belong to a critical droplet, and therefore the critical density, Eq. (\ref{vcgen1}), depends on $m-1$ iterations of the $\ln$ function. More specifically, we can write the probability to expand a hyper-cube up to linear size $L$ as
\begin{align}
P_{m}=\prod^{L}_{l=1}f^{d}_{m-2,m}(l) ,
\end{align}
where $f_{n,m}(l)$ is defined by the recursion relation
\begin{align}
&f_{n+1,m}(l)=1-\left(1-\prod^{l}_{k=1}f_{n,m}(k)\right)^{(m-1)l^{d+n-1}} ,\nonumber\\
&f_{0,m}(l)=\beta_{m}(l)=\nonumber\\
&=\frac{1-\rho^{(m-1)l}+\sqrt{\left(1+\rho^{(m-1)l}\right)^{2}-4\rho^{ml}}}{2} .
\end{align}
Approximating $P_{m}$ as was done before (changing the products to integrals of $\ln$ functions, etc.), and solving the equation $1=L^{d}P_{m}$ yields Eq. (\ref{vcgen1}). 

If the number of effective dimensions of the hyper-rhomboid is at least $m$, the system is either almost completely frozen or almost completely unfrozen and there is no division to sections or clusters. Assuming that there are $s$ small sides, then in the integrals of $g_{m-1}$ (as in Eqs. (\ref{eqvgen}) and (\ref{eqvgen3})) the upper limits of the first $s$ integrals can be taken to zero and the lower limits of the last $d-s-1$ integrals can be taken to infinity, such that only one integral remains, which yields $\lambda_{d-s,m}/\left(vV_{s}\right)^{1/(d-s-m+1)}$. The other outer integrals yield unimportant constants (similar to $\mu$ in the $m=3$ case), and each iteration of the $f$ function gives another exponent, such that in the end the critical density is
\begin{align}
v_{c}=\frac{1}{V_{s}}\left(\frac{\lambda_{d-s,m}}{\ln_{(m-1)}V^{1/(d-s)}}\right)^{d-s-m+1} ,
\end{align}
which is similar to the behavior of a $d-s$ dimensional system.

If $d_{eff}<m$, the hyper-rhomboid is divided into clusters. If $d_{eff}=m-1$, the characteristic size of the clusters is determined by the characteristic size of the clusters on the $d-1$ dimensional hyper-surfaces, which ultimately depends on the size of the subsection on the hyper-surface unblocked via the $m=2$ process, which is now a tunnel. The characteristic cluster size is thus $\rho^{-cV_{d-m+1}}$ in this case, where $c$ is some constant which depend on $d$ and $m$. If $d_{eff}=m-2$, the characteristic size of the clusters is ultimately determined by the size of the subsections on the hyper-surface (which is a tunnel) unblocked via the $m=3$ process, which is $\sim\exp\left[V_{d-m+2}\exp(-2\lambda_{d,m}/v)\right]$. In general, the characteristic cluster size is determined by the size of the subsections on the hyper-surface (which is a tunnel) unblocked via the $m-d_{eff}+1$ process, which is $\sim\exp^{d_{eff}-1}\left[V_{t}\exp(-2\lambda_{d,m}/v)\right]$, with $\exp^{n}$ being an exponential iterated $n$ times. Due to the combination of the iterated $\ln$ functions and the exponents, the critical vacancy density in all cases scales as
\begin{align}
v_{c}\sim\left[\ln_{(m-2)}V^{1/(d-t)}_{s}\right]^{-d+t+m-1} ,
\end{align}
which is similar to a $d-t-1$ dimensional system in a $m-1$ model.

\section{The relation between the correlation between frozen sites and the fluctuations of $n_{PF}$}
\label{snpf}

Consider the correlation between sites $\vec{i}$ and $\vec{j}$
\begin{align}
c_{\vec{i},\vec{j}}=\frac{1}{A}\sum_{\alpha}\left(n^{\alpha}_{\vec{i}}-\rho n_{PF}\right)\left(n^{\alpha}_{\vec{j}}-\rho n_{PF}\right) ,\label{ap1}
\end{align}
where the sum is over all configurations $\alpha$, the total number of configurations is $A$, for each site $n^{\alpha}_{\vec{i}}$ is equal to one if the site contains a frozen particle in configuration $\alpha$ and is equal to zero in all other cases, and $\rho n_{PF}$ is the average fraction of frozen sites
\begin{align}
\rho n_{PF}=\frac{1}{AV}\sum_{\vec{i},\alpha}n^{\alpha}_{\vec{i}} .\label{ap2}
\end{align}
The average correlation is
\begin{align}
C_{av}=\frac{1}{V^{2}}\sum_{\vec{i},\vec{j}}c_{\vec{i},\vec{j}} ,\label{ap3}
\end{align}
where $V$ is the volume of the system. Using the definition of $c_{\vec{i},\vec{j}}$, Eq. (\ref{ap1}), in Eq. (\ref{ap3}) yields
\begin{align}
&C_{av}=\frac{1}{AV^{2}}\sum_{\vec{i},\vec{j},\alpha}\left(n^{\alpha}_{\vec{i}}-\rho n_{PF}\right)\left(n^{\alpha}_{\vec{j}}-\rho n_{PF}\right)=\nonumber\\
&=\frac{1}{AV^{2}}\sum_{\vec{i},\vec{j},\alpha}\left[n^{\alpha}_{\vec{i}}n^{\alpha}_{\vec{j}}-n^{\alpha}_{\vec{i}}\rho n_{PF}-n^{\alpha}_{\vec{j}}\rho n_{PF}+\rho^{2}n^{2}_{PF}\right] .
\end{align}
In the second term we sum over $\vec{j}$, in the third term we sum over $\vec{i}$, and in the fourth term we sum over $\vec{i},\vec{j}$ and $\alpha$, such that
\begin{align}
&C_{av}=\frac{1}{AV^{2}}\sum_{\vec{i},\vec{j},\alpha}n^{\alpha}_{\vec{i}}n^{\alpha}_{\vec{j}}-\frac{1}{AV}\sum_{\vec{i},\alpha}n^{\alpha}_{i}\rho n_{PF}-\nonumber\\
&-\frac{1}{AV}\sum_{\vec{j},\alpha}n^{\alpha}_{j}\rho n_{PF}+\rho^{2}n^{2}_{PF} .
\end{align}
Using Eq. (\ref{ap2}) in the second and third term yields
\begin{align}
C_{av}=\frac{1}{AV^{2}}\sum_{\vec{i},\vec{j},\alpha}n^{\alpha}_{\vec{i}}n^{\alpha}_{\vec{j}}-\rho^{2}n^{2}_{PF} .\label{apcav}
\end{align}

Now consider the fluctuations in $\rho n_{PF}$
\begin{align}
\sigma^{2}=\frac{1}{A}\sum_{\alpha}\left(\rho n^{\alpha}_{PF}\right)^{2}-\rho^{2}n^{2}_{PF} ,\label{ap4}
\end{align}
where $\rho n^{\alpha}_{PF}$ is the fraction of frozen sites in configuration $\alpha$
\begin{align}
\rho n^{\alpha}_{PF}=\frac{1}{V}\sum_{\vec{i}}n^{\alpha}_{\vec{i}} .\label{ap5}
\end{align}
Using Eq. (\ref{ap5}) in Eq. (\ref{ap4}) yields
\begin{align}
&\sigma^{2}=\frac{1}{A}\sum_{\alpha}\left(\frac{1}{V}\sum_{\vec{i}}n^{\alpha}_{\vec{i}}\right)\left(\frac{1}{V}\sum_{\vec{j}}n^{\alpha}_{\vec{j}}\right)-\rho^{2}n^{2}_{PF} ,
\end{align}
which is equal to Eq. (\ref{apcav}), and thus $\sigma^{2}=C_{av}$.

Note that this result is valid in more general conditions: $n^{\alpha}_{\vec{i}}$ can receive values other than zero or one, the different configurations can have different weights, and the system can be continuous.

\end{document}